\documentstyle[12pt,epsfig]{report}
\large
\newcommand{\beq}{\begin{equation}}
\newcommand{\eeq}{\end{equation}}
\newcommand{\bra}{\begin{array}}
\newcommand{\era}{\end{array}}
\newcommand{\be}{\beta}

\let\include\input
\def\be{\begin{equation}}
\def\ee{\end{equation}}
\def\bea{\begin{eqnarray}}
\def\eea{\end{eqnarray}}

\def\[{\bigl[}
\def\]{\bigr]}
\def\({\bigl(}
\def\){\bigr)}

\begin{document}
\makeatletter \@addtoreset{equation}{section}
\renewcommand{\theequation}{\thesection.\arabic{equation}}
\title{\rightline{\mbox{\small
{2011-2012}}} \textbf{ Sym\'etrie en Physique:  Alg\`ebres de Lie, Th\'eorie des groupes et Repr\'esentations } \\
\textbf{ }}
\author{ \textbf{Adil Belhaj}\thanks{belhaj@unizar.es} \\
{\small   Centre of Physics and Mathematics, CPM-CNESTEN, Rabat, Morocco}\\
{\small }} \maketitle
 \vspace{-10cm}
\begin{abstract}
These notes form an introduction  to Lie algebras  and group theory.
Most of the material  can be found in  many works by various authors
given in the list of references. The reader is referred to such
works for more detail.

\end{abstract}


\thispagestyle{empty} \newpage \setcounter{page}{1} \newpage
 Ce travail  est une note   de cours sur  la sym\'etrie en physique:
 Alg\`ebres de Lie et la th\'eorie des
 groupes que nous avons  pr\'epar\'es  pour les
    \'etudiants  de DESA et  Master Physique Math\'ematique:   Lab de Physique des Hautes Energies: Mod\'elisation et
Simulation, Facult\'e des Sciences,  Universit\'e Mohammed V
Agdal-Rabat, Maroc; et \'egalement pour  les doctorants en physique
th\'eorique de l'Universit\'e de Zaragoza,   Espagne. Ce sont des
notes pr\'eliminaires d\'estin\'ees pour des jeunes chercheurs.
\begin{center}
Remerciement:
 \end{center}
Je tiens \`a exprimer mes profonds remerciements \`a ma ch\`ere
m\`ere pour le soutien moral et le support social  sans \'egal.

\tableofcontents

\chapter{   Introduction G\'en\'erale}
\pagestyle{myheadings}
\markboth{\underline{\centerline{\textit{\small{ Introduction
G\'en\'erale}}}}}{\underline{\centerline{\textit{\small{
Introduction G\'en\'erale}}}}} Les alg\`{e}bres de Lie et la
th\'eorie des groupes sont largement utilis\'ees dans de nombreuses
branches de la physique et de la chimie. Actuellement,  elles
peuvent \^etre consid\'er\'ees comme une composante essentielle dans
l'\'etude de la physique des hautes \'energies,  la physique
math\'ematique et la mati\`eres condens\'ee. En particulier, elles
apparaissent dans la physique des particules et la construction de
la th\'eorie de jauge \`a  quatre dimensions, \`a  partir de la
th\'eorie des cordes et des th\'eories M et F. Rappelons que la
th\'eorie des cordes est l'une des voies envisag\'ees pour r\'egler
une des questions majeures de la physique des  hautes \'energies:
fournir une description de la gravit\'e quantique c'est-\`a-dire
l'unification de la m\'ecanique quantique  et de la th\'eorie de la
relativit\'e g\'en\'erale. Cependant, il existe en fait cinq
th\'eories des supercordes. Elles ont en commun un univers \`a 10
dimensions qui ne poss\`ede pas de tachyons et supposent l'existence
d'une supersym\'etrie sur la surface d'univers des supercordes. Ces
mod\`eles sont class\'es comme suit:
\begin{enumerate}
\item
 Type I: supercordes ouvertes ou ferm\'ees, avec le groupe de sym\'etrie SO(32)
 \item Type IIA: supercordes ferm\'ees uniquement, non-chiralit\'e
 \item Type IIB: supercordes ferm\'ees uniquement, chiralit\'e
 \item Type HO: supercordes ferm\'ees uniquement, h\'et\'erotique avec le groupe de sym\'etrie SO(32)
\item Type HE: supercordes ferm\'ees uniquement, h\'et\'erotique, avec  le groupe de sym\'etrie $E_8\times E_8$.
\end{enumerate}

A partir de cette classification, nous pouvons consid\'erer
diff\'erents exemples des alg\`{e}bres de Lie. Néanmoins, l'absence
de telles sym\'etries dans certains mod\`eles de supercordes peuvent
\^etre rem\'edi\'ee en faisant le lien entre la g\'eom\'etrie et les
alg\`ebres de Lie, en envisageant pluiseurs  sc\'enarios de
compactification.

Ces notes  offrent une introduction aux alg\`{e}bres de Lie,  la
th\'eorie des groupes, ainsi que ses repr\'esentations. Ils
comprennent de nombreux exemples et exercices. Ils sont destin\'es
aux \'etudiants en physique des hautes energies, la  physique
math\'ematique et la physique de la mati\`ere condens\'ee.

Le plan de ces  notes  est  comme suit:

Le deuxi\`eme  chapitre est une introduction sur les ag\`ebres de
Lie. Cette partie pr\'esente les notions de base. En particulier,
nous introduisons une partie traitant la classification des
alg\`ebres de type ADE.

Le troisi\`eme chapitre est consacr\'e essentiellement \`a
 introduire quelques aspects de la sym\'etrie  $sl(2,C)$ et ses  repr\'esentations.
Pour ce faire, nous commen\c{c}ons par introduire  $sl(2,C)$  et
nous pr\'esentons dans la suite de ce chapitre une mani\`ere de
construire  ses repr\'esentations.

Dans le quatri\`eme chapitre on \'etudie  les sous alg\`ebres de
Cartan et les syst\`emes de racines.

Le cinqui\`eme chapitre est consacr\'e  \`a  l'\'etude  des
syst\`emes des racines. Nons donnons  la  classification  ADE.

 Dans le  sixi\'eme chapitre, nous \'etudions les matrices de
 Cartan  et  nous donnons \'egalement la classification des diagrammes de Dynkin.

 Le septi\`eme chapitre  pr\'esente quelques g\'en\'eralit\'es sur  la
 th\'eorie des groupes.

 Dans le huiti\`eme chapitre, nous \'etudions les repr\'esentations des groupes.

 Le dernier chapitre est consacr\'e \`a  une discussion des repr\'esentations des  groupes
 de Lie.

\chapter{   Notions sur les alg\`{e}bres de Lie}
\pagestyle{myheadings}
\markboth{\underline{\centerline{\textit{\small{ Notions sur les
alg\`{e}bres de Lie}}}}}{\underline{\centerline{\textit{\small{
Notions sur les alg\`{e}bres de Lie}}}}}

Dans ce chapitre, nous allons pr\'esenter quelques notions sur les
alg\`{e}bres de Lie. En premier temps, nous donnons des
d\'{e}finitions  et   des exemples simples. On termine ce chapitre
par une   classification des alg\`{e}bres de Lie de type  ADE.

\section{D\'{e}finitions et exemples}

\subsection{D\'{e}finitions}
Une alg\`{e}bre de Lie $g$ est un espace vectoriel sur un corps $C$
(o\`u $R$)  muni de l'op\'{e}ration $\left[ ,\right] $:
 \begin{eqnarray*}
 \left[ ,\right]: \quad  g.g &\rightarrow &g\\
 (x,y) &\rightarrow& \left[ x,y\right]=xy-yx
\end{eqnarray*}  v\'erifiant   les trois conditions suivantes:
\\ 1. lin\'{e}arit\'{e}
\begin{eqnarray*}
\left[ \alpha x+\beta y,z\right] =\alpha \left[ x,z\right] +\beta \left[ y,z%
\right],\qquad \alpha, \beta \in C \end{eqnarray*}
 2. antisym\'{e}trie
\begin{eqnarray*}
\left[ x,y\right] =-\left[ y,x\right]
\end{eqnarray*} 3.
identit\'{e} de Jacobi
\begin{eqnarray*}
\left[ \left[ x,y\right],z\right] +\left[ \left[ y,z\right],x\right] +%
\left[ \left[ z,x\right],y\right] =0 \end{eqnarray*} pour tous $x$,
$y$, $z \in g$. \subparagraph{Remarques:}
\begin{enumerate} \item L'op\'{e}ration
$\left[ ,\right] $ est appel\'{e}e le crochet de Lie o\`u bien le
commutateur de Lie.
\item La dimension d'une alg\`ebre de Lie $g$ est la  dimension en tant qu'espace
vectoriel.
\item Lorsque  la dimension de l'alg\`ebre de Lie $g$ est finie, on peut construire une base.
 \end{enumerate}
\subsection{Relations entre les g\'{e}n\'{e}rateurs d'une alg\'{e}bre de Lie }
Soit $g$ une alg\`{e}bre de Lie ayant une base $\{
t_{a},\;a=1,\ldots, \dim g \}$. La  structure d'alg\`{e}bre de Lie
est donn\'{e}e  aussi par des relations de commutation entre les
g\'{e}n\'{e}rateurs $ t_{a}$ comme suit
\begin{eqnarray*}
\left[ t^{a},t^{b}\right] =c_{c}^{ab}t^{c},\;\; \qquad a,b,c=1,\ldots,   \dim g
\end{eqnarray*}
o\`u $c_{c}^{ab}$ sont des tenseurs antisym\'{e}triques constantes
\begin{eqnarray*}
c_{c}^{ab}=-c_{c}^{ba}.
\end{eqnarray*}
\subparagraph{Remarques:}\begin{enumerate}
\item  Les coefficients $c_{c}^{ab}$ sont appel\'es constantes de structure.
\item Si tous les
crochets de Lie sont nuls, l'alg\`{e}bre est
 dite ab\'elienne.
\item  Si deux alg\`{e}bres de dimension finie ont les m\^{e}mes
relations de commutation,  elles sont isomorphes.
 \end{enumerate}
\subsection{Exemples}
Donnons quelques exemples d'alg\`ebres de Lie que nous rencontrons
en physique et  nous les exprimons en fonction de leurs constantes
de structure.\\
{\bf Exemple 1:} Alg\`{e}bre de Heisenberg.\\
Notre premier exemple est l'alg\`{e}bre de Heisenberg qui  apparait
dans les relations de la quantification cannonique de la
m\'{e}canique quantique. Elle est engendr\'{e}e
 par le syst\`{e}me des g\'{e}n\'{e}rateurs $\{a^{+},a^{-},I\}$. Sa structure est donn\'{e}e par
\begin{eqnarray*}
\lbrack a^{+},a^{-}]=I,\qquad  \lbrack a^{\pm },a^{\pm }]=0,\qquad
\lbrack a^{\pm },I]=0.
\end{eqnarray*}
{\bf Exemple 2:}  Alg\`{e}bre  de Lie $so(3)$.  On consid\`{e}re
l'alg\`{e}bre
 des rotations \`{a} 3 dimensions. La structure de $so(3)$ est donn\'{e}e par
les relations de commutation suivantes
 \begin{eqnarray*}
\left[ t^{i},t^{j}\right] =-i\varepsilon ^{ijk}t_{k},
\end{eqnarray*}
o\`u $\varepsilon ^{ijk}$ est le tenseur de levi-civita
compl\`etement antisym\'{e}trique
$\varepsilon^{ijk}=-\varepsilon^{jik}, \;\varepsilon ^{iik}=0,\;
\varepsilon ^{ijk}=1, \;i\neq j\neq
k$.\\
{\bf Exemple 3:} L'ensemble $M_{2}(C)$ des matrices d'ordre 2, muni de la multiplication matricielle,
\begin{eqnarray*}
M= \left(
\begin{array}{cc}
\alpha & \beta \\
\gamma & \sigma%
\end{array}%
\right)= \alpha\left(
\begin{array}{cc}
1 & 0 \\
0 & 0%
\end{array}%
\right)+\beta\left(
\begin{array}{cc}
0 & 1 \\
0 & 0%
\end{array}%
\right)+\gamma\left(
\begin{array}{cc}
0 & 0 \\
1& 0%
\end{array}%
\right)+\sigma\left(
\begin{array}{cc}
0 & 0\\
0 & 1%
\end{array}%
\right)
\end{eqnarray*}
 est  consid\'{e}r\'{e} comme une  alg\`{e}bre de Lie
de dimension 4 complexes.

\section{Sous alg\`{e}bres de Lie}
\subsection{D\'{e}finitions}
\subsubsection{D\'{e}finition 1:} Soit $g_{0}$ un sous espace vectoriel d'une alg\`{e}bre de Lie
$g$. $g_{0}$ est dite une sous alg\`{e}bre de Lie de $g$ si et
seulement si
\begin{eqnarray*}
\left[ g_{0}, g_{0}\right] \subset g_{0}.
\end{eqnarray*}
Autrement,
\begin{eqnarray*}
\forall, x \in   g_{0},\; et \;y \in  g_{0}\; alors\; \left[
x,y\right] \in g_{0}.
\end{eqnarray*}
\subsubsection{D\'{e}finition 2:}
Soit $I\subset g$ (sous alg\`{e}bre de Lie de $g$).  $I$ est dite un
id\'{e}al de $g$ si et seulement si
\begin{eqnarray*}
 \left[ I,g\right] \subset I.
\end{eqnarray*}
Autrement,
\begin{eqnarray*}
\forall x \in  I ,\;\; y \in g \; alors\; \left[ x,y\right] \in I.
\end{eqnarray*}
\subsection{Exemples}
Nous  pr\'esentons quelques exemples.\\
\\
{\bf  Exemple 1}: $\left\{ 0\right\} $ est un id\'{e}al de $g$
(id\'{e}al
trivial).\\
{\bf  Exemple 2}: L'alg\`ebre de Lie $g$ est elle-m\^eme un id\'eal.
En effet, on a
\begin{eqnarray*}
\left[ g,g\right] \subset g.
\end{eqnarray*}
{\bf  Exemple 3}: Le centre d'une alg\`{e}bre de Lie.\\
Le centre de $g$ not\'e $Z(g)$ est d\'{e}fini par
\begin{eqnarray*}
Z(g)=\{x, \;\;\left[ x,y\right] =0,\;\;\forall y \in  g\}.
\end{eqnarray*}
 $Z(g)$ est un id\'{e}al ab\'{e}lien de $g$.\\
{\bf D\'emonstration}:\\ Pour d\'emontrer  ca,  il faut
 v\'erifier que si $x \in Z(g)$ alors $\left[ x,y\right] $ $\in $ $%
Z(g)$, pour tout $y\in g$. En effet, nous avons
\begin{eqnarray*}
x\in z(g) &\; alors\; &\left[ x,y\right] =0 \\
&\; \;&\left[ x,x\right] \subset Z(g).
\end{eqnarray*}
{\bf  Exemple 4}:  D\'erivation d'une alg\`ebre de Lie.\\
Soit $g$ une alg\`{e}bre de Lie. On appelle alg\`ebre de Lie d\'eriv\'ee de $g$ l'ensemble d\'efini par  $g^{\prime }=\left[ g,g\right] $.\\
 $g^{\prime }$ est un id\'{e}al $\left[ g^{\prime },g\right] \subset g^{\prime
 }$. \\
{\bf D\'emonstration}:\\Puisque on a $\left[ g^{\prime },g\right] =\left[ \left[ g,g\right] ,g\right] \subset %
\left[ g,g\right] \subset g^{\prime }$.

\section{Alg\`{e}bres quotients}
Soit $g$ une alg\`{e}bre de Lie. Il est naturel de factoriser par un
id\'{e}al pour construire  un espace quotient de $g$  par un
id\'{e}al $I$. L'espace quotient  est d\'{e}fini par
\begin{eqnarray*}
\ g_{ I}=\{\dot{x},\;\dot{x}=x+\mu,\; x \in  g,\mu  \in  I\}.
\end{eqnarray*}
Notons que $\ g_{I}$  poss\`ede aussi  une structure d'alg\`ebre de
Lie
\begin{eqnarray*} \ \left[ \ g_{I},\ g_{
I}\right] \subset \ g_{ I}.
\end{eqnarray*}
Calculons le commutateur  $[\dot{x},\dot{y} ] $. En effet, nous avons
\begin{eqnarray*}
\ \left[ \dot{x}, \dot{y}\right]  &=&\left[ x+\mu,
y+\mu \right]  \\
&=&\left[ x,y\right] +I\subset g_{I}.
\end{eqnarray*}

\section{Homomorphismes d'une alg\`{e}bre de Lie}
\subsection{D\'efinitions}
Rappelons que la loi d'une alg\`{e}bre de Lie est abstraite.  En pratique, on a besoin des r%
\'{e}alisations (repr\'{e}sentations) pour \ l'exploiter dans des
applications physiques.  En g\'{e}n\'{e}rale, on peut parler des repr\'{e}sentations
matricielles ou bien des r\'{e}alisations   par
des op\'{e}rateurs diff\'{e}rentiels agissant  sur des espaces
vectoriels.  Ce passage est assur\'{e} par un homomorphisme de
l'alg\`{e}bre.
\subsubsection{D\'{e}finition 1}

Soient $g$ et $g^{\prime }$ deux alg\`{e}bres de Lie. Un homomorphisme (repr\`{e}sentation) $\phi $ d'une alg\`{e}bre de Lie est d\'{e}fini par:
\begin{eqnarray*}
\phi :&g&\rightarrow g^{\prime }\\
&x&\rightarrow \phi (x),
\end{eqnarray*}
v\'erifiant les conditions suivantes
\begin{eqnarray*}
 \phi \left( \alpha x+\beta y\right) &=&\alpha \phi (x)+\beta \phi
(y) \\
 \phi \left( \left[ x,y\right] \right) &=&\left[ \phi (x),\phi
(y)\right].
 \end{eqnarray*}
pour tous  $x$ et  $y$ dans $g$.  $\alpha$ et  $\beta$  sont des
complexes.
\subsubsection{D\'{e}finition}
Une repr\'esentation de $g$ est un homomorphisme   d'alg\`ebre de
Lie de $g$  dans $gl(V )$  o\`u  $gl(V )$  est l'alg\`ebre de Lie
associ\'ee au groupe des automorphismes $GL(V )$,   o\`u   $V$ est un
espace vectoriel:
\begin{eqnarray*}
\phi :&g&\rightarrow gl(V)\\
&x&\rightarrow \phi (x)= \mbox{ matrice, OP}.
\end{eqnarray*}
o\`u OP  est un op\'{e}rateur diff\'{e}rentiel.\\ {\bf Remarque}: Si
l'espace vectoriel V est de dimension $n$, la repr\'esentation $g
\to gl(V )$ est dite une repr\'esentation de dimension $n$.
\subsubsection{Exemple}
On reprend l'exemple de l'alg\`ebre de Lie de Heisenberg. Rappelons
qu'on  d\'efinit l'alg\`ebre de Heisenberg  comme suit:
\begin{eqnarray*}
\left[ a^{+},a^{-}\right] =I.
\end{eqnarray*}
Dans le plan complexe param\'{e}trise par $z$, une r\'ealisation est
donn\'ee par
\begin{eqnarray*}
 \phi (a^{+})&=&z\\
\phi (a^{-})&=-&\frac{\partial }{\partial z}\\
 \phi (I)&=&1.
\end{eqnarray*}
En effet,  on peut v\'erifier facilement que
\begin{eqnarray*}
\left[ \phi (a^{+}),\phi (a^{-})\right]  &=&\phi (a^{+})\phi
(a^{-})-\phi
(a^{-})\phi (a^{+}) \\
&=-&z\frac{\partial }{\partial z}+\frac{\partial }{\partial z}z \\
&=&-0+1 \\
&=&\phi (I).
\end{eqnarray*}
\subsection{ Repr\'esentation adjointe}
Il existe une repr\'esentation donn\'ee par les constantes de
structure d'une alg\`ebre de Lie. On l'appelle la repr\'esentation
adjointe. Elle  est d\'efinie par la relation suivante
\begin{eqnarray*}
 ad:   g  \rightarrow gl(g)\\
 x \rightarrow [x, . ]
\end{eqnarray*}
   o\`u  $ad_x.y=ad_x(y)=[x , y ]$.\\
 {\bf  Proposition}:  La dimension de la repr\'esentation adjointe
est la dimension de l'espace vectoriel sur lequel elle agit.  Sa dimension  coincide avec celle de l'ag\`ebre $g$.

\section{Alg\`{e}bre de Lie des matrices}
L'alg\`{e}bre des matrices $n\times n$  muni de la multiplication
matricielle est une alg\`{e}bre de Lie de dimension $n^{2}$. On la
note $gl(n,C)$.  Par la suite, on  note   un \'{e}l\'{e}ment de  $gl(n,{C})$  par
\begin{eqnarray*}
X=(x_{ij}), \qquad i,j=1,\ldots, n.
\end{eqnarray*}
Dans le reste de cette section, nous  \'{e}tudions les sous alg\`{e}bres de $gl(n,C)$.

\subsection{Alg\`ebre $sl(n,C)$}
Notre premiere sous  alg\`ebre  de $gl(n,C)$  exemple est l'ensemble des matrices de  trace nulle.
Elle est d\'efinie par
\begin{eqnarray*}
sl(n,C)=\left\{ X\in gl(n,C),\;\;\; trX=0\right\}.
\end{eqnarray*}
{\bf  Remarque}: la  dimension  de $sl(n,C)$ est $n^{2}-1$.\\
\\
{\bf Exemple}: Prenons l'exemple de l'alg\`{e}bre $sl(2,C)$  qui est
  l'ensemble des matrices $2\times 2$ avec la trace nulle
\begin{eqnarray*}
sl(2,C)=\left\{ X \in  gl(2,C), \;\;\;trX=0\right\}.
\end{eqnarray*}
Si $X$ est une matrice de $gl(2)$, alors elle peut s'\'ecrire comme
suit
\begin{eqnarray*}
\ \ \ \ \ X=\left(
\begin{array}{cc}
a & b \\
c & d%
\end{array}%
\right).
\end{eqnarray*}
La condition  $trX=0$ implique que $a+d=0$. On  d\'eduit que $a=-d$.
Cela nous permet de trouver
\begin{eqnarray*}
\ \ \ \ \ X=\left(
\begin{array}{cc}
a & b \\
c & -a%
\end{array}%
\right).
\end{eqnarray*}
Cette matrice  peut s'\'ecrire \'egalement   sous la forme suivante
\begin{eqnarray*}
 X=\left(
\begin{array}{cc}
a & b \\
c & -a%
\end{array}%
\right) =a\left(
\begin{array}{cc}
1 & 0 \\
0 & -1%
\end{array}%
\right) +b\left(
\begin{array}{cc}
0 & 1 \\
0 & 0%
\end{array}%
\right) +c\left(
\begin{array}{cc}
0 & 0 \\
1 & 0%
\end{array}%
\right)
\end{eqnarray*}
 o\`u $a,b,c\in C$. On  d\'eduit que la dimension de $sl(2,C)$ est 3  complexes $(3=2^{2}-1)$.\\
Si  on  travaille dans la base
\begin{eqnarray*}
J_{0}=\left(
\begin{array}{cc}
1 & 0 \\
0 & -1%
\end{array}%
\right), \qquad
 J_{+}=\left(\begin{array}{cc}
0 & 1 \\
0 & 0%
\end{array}%
\right), \qquad  J_{-}=\left(
\begin{array}{cc}
0 & 0 \\
1 & 0%
\end{array}%
\right),
\end{eqnarray*}
on peut   calculer les relations de commutation de $sl(2,C)$ comme suit
\begin{eqnarray*}
\left[ J_{0},J_{+}\right]& =&2J_{+}\\
 \left[ J_{0},J_{-}\right]
&=&-2J_{-}\\
\ \left[ J_{+},J_{-}\right] &=&J_{0}.
\end{eqnarray*}

\subsection{Alg\`ebre $su(n,C)$}
Le deuxi\`eme exemple sera  les  matrices $n\times n$
anti-herm\'{e}tique de trace nulle
\begin{eqnarray*}
 su(n)=\left\{ X \in gl(n,C),\;\;
X+X^{+}=0,\:\;trX=0\right\}.
\end{eqnarray*}
La  dimension  de  $su(n,C)$ est   $n^{2}-1$,  ($\dim su(n)=n^{2}-1$).\\
\\
{\bf Exemple}:  L'alg\`{e}bre de Lie $su(2)$ \\
$su(2)$ est definit par
\begin{eqnarray*}
\left\{ X \in gl(2,C), \;\;X+X^{+}=0 ,\;\;  trX=0\right\}
\end{eqnarray*}
qui est  un espace vectoriel (de $\dim $ 3) des matrices complexes
anti-hem\'{e}tiques $2\times 2$ de trace nulle. En effet,
consid\'erons  un \'el\'ement  $X$   de $gl(2,C)$
\begin{eqnarray*}
 X=\left(
\begin{array}{cc}
a & b \\
c & d%
\end{array}%
\right).
\end{eqnarray*}
Les contraintes sur la matrice  $X$  sont
\begin{eqnarray*}
a+d=0\\
 a=-\overline{a}, d=-\overline{d}\\
b=-\overline{c} \\
 c=-\overline{b}
\end{eqnarray*}
La salution de ces constraintes est donn\'ee par les relations
suivantes
\begin{eqnarray*}
a=ix_{3}\\
 b=x_{2}+ix_{1} \\
 c=-x_{2}+ix_{1}
\end{eqnarray*}
o\`u  $x_{1},x_{2}$ et $x_{3}\in  R$.  La matrice $X$  peut
s'\'ecrire comme
\begin{eqnarray*}
 X=\left(
\begin{array}{cc}
ix_{3} & x_{2}+ix_{1} \\
-x_{2}+ix_{1} & -ix_{3}%
\end{array}%
\right).
\end{eqnarray*}
Nous observons alors que la dimension de   $su(2)$ est 3 r\'eel.  Une base de
$su(2)$ est  donn\'ee par
\begin{eqnarray*}
I=\left(
\begin{array}{cc}
0 & 1 \\
-1 & 0%
\end{array}%
\right), \qquad  J=\left(
\begin{array}{cc}
0 & i \\
i & 0%
\end{array}%
\right),\qquad K=\left(
\begin{array}{cc}
i & 0 \\
0 & -i%
\end{array}%
\right).
\end{eqnarray*}
On constate  que les relations de commutation de $su(2)$  sont comme suit
\begin{eqnarray*}
 \left[ I,J\right] &=&IJ-JI=2K\\
\left[ J,K\right] &=&JK-KJ=2I\\
\left[ K,I\right] &=&KI-IK=2J.
\end{eqnarray*}
En introduisant les  matrices de Pauli suivantes
\begin{eqnarray*}
\sigma _{1}=\left(
\begin{array}{cc}
0 & 1 \\
1 & 0%
\end{array}%
\right), \qquad \sigma _{2}=\ \left(
\begin{array}{cc}
0 & -i \\
i & 0%
\end{array}%
\right),\qquad \sigma _{3}=\left(
\begin{array}{cc}
1 & 0 \\
0 & -1%
\end{array}%
\right),
\end{eqnarray*}
les \'el\'ements de base sont alors de la forme suivante
\begin{eqnarray*}
I=-i\sigma _{2},\qquad  I=i\sigma _{1},\qquad  K=i\sigma _{3}.
\end{eqnarray*}

\subsection{Alg\`{e}bres de Lie $so(n)$ et $sp(n)$}
Soit $J$ une matrice quelconque.  On peut d\'{e}finir une s\'{e}rie
de sym\'{e}tries de Lie \`{a} travers la relation suivante
\begin{eqnarray*}
\ g_{J}=\left\{ X\in sl(2n,C), \;\; JX+X^{t}J=0\right\}.
\end{eqnarray*}
{\bf Proposition:}
$g_{J}$ est une sous alg\`{e}bre de Lie de  $sl(2n,C)$.\\
Prenons  un choix particulier de $J$  tel que
\begin{eqnarray*}
 J=\left(
\begin{array}{cc}
0 & I_{n\times n} \\
I_{n\times n} & 0%
\end{array}%
\right) _{2n\times 2n}
\end{eqnarray*}
Dans ce cas,  l'alg\`{e}bre de Lie $g_{J}$   devient $so(2n)$. Elle
est d\'efinie par:
\begin{eqnarray*}
so(2n)=\left\{ X\in sl(2n), \;\; \left(
\begin{array}{cc}
0 & I_{n\times n} \\
I_{n\times n} & 0%
\end{array}%
\right) X+X^{t}\left(
\begin{array}{cc}
0 & I_{n\times n} \\
I_{n\times n} & 0%
\end{array}%
\right) =0_{2n\times 2n}\right\}.
\end{eqnarray*}
Int\'eressons-nous plus particuli\`erement au cas o\`u  $n = 1$.
Dans ce cas, la  matrice $J$ est donn\'{e}e par
\begin{eqnarray*}\
J=\left(
\begin{array}{cc}
0 & 1 \\
1 & 0%
\end{array}%
\right).
\end{eqnarray*}
Soit  un \'el\'ement  $X$   de $sl(2,C)$ s'\'ecrivant sous la forme
suivante
\begin{eqnarray*}
X=\left(
\begin{array}{cc}
x & y \\
z & -x%
\end{array}%
\right).
\end{eqnarray*}  Consid\'erons la base canonique form\'ee par les
matrices suivantes \begin{eqnarray*}
 E_{11}=\left(
\begin{array}{cc}
1 & 0 \\
0 & 0%
\end{array}%
\right),\qquad E_{12}=\left(
\begin{array}{cc}
0 & 1 \\
0 & 0%
\end{array}%
\right), \qquad   E_{21}=\left(
\begin{array}{cc}
0 & 0 \\
1 & 0%
\end{array}%
\right),\qquad  E_{22}=\left(
\begin{array}{cc}
0 & 0 \\
0 & 1%
\end{array}%
\right).
\end{eqnarray*}
 En terme de  $E_{ij}$,  un  \'el\'element $X$   dans $ sl(2,C)$ s'exprime  comme
   \begin{eqnarray*} X=x\left(
\ E_{11}-E_{22}\right) +yE_{12}+zE_{21}.\end{eqnarray*} Cherchons  la
solution de la  contrainte  $JX+X^{T}J=0_{2\times 2}$. En effet, nous
calculons l'expression
\begin{eqnarray*}
\ JX+X^{T}J &=&\ \left(
\begin{array}{cc}
0 & 1 \\
1 & 0%
\end{array}%
\right) \left(
\begin{array}{cc}
x & y \\
z & -x%
\end{array}%
\right) +\left(
\begin{array}{cc}
x & z \\
y & -x%
\end{array}%
\right) \left(
\begin{array}{cc}
0 & 1 \\
1 & 0%
\end{array}%
\right)  \\
&=&\left(
\begin{array}{cc}
z & -x \\
x & y%
\end{array}%
\right) +\left(
\begin{array}{cc}
z & x \\
-x & y%
\end{array}%
\right).
\end{eqnarray*}
La condition $JX+X^{T}J=0_{2\times 2}$ exige que
\begin{eqnarray*}
 y=0, \qquad
 z=0,\qquad\;\; et\;\;   x\;est\; quelconque.
 \end{eqnarray*}
Si $X$ est une matrice de $so(2)$, alors elle peut s'\'ecrire  comme
suit
\begin{eqnarray*}
X=x(E_{11}-E_{22})=x\left(
\begin{array}{cc}
1 & 0 \\
0 & -1%
\end{array}%
\right) \ \ avec\ \ \ x\in R. \end{eqnarray*}

\subsubsection{Remarques:}
\begin{itemize}
\item  Elle  est une sous alg\`{e}bre de Lie de $sl(2,R)$.
\item Sa dimension  est   $ dim\; so(2)=1$.  Par cons\'equent,   elle est
ab\'{e}lienne.
\item  Pour le  cas g\'{e}n\'{e}rale $n>1$,
l'alg\`{e}bre de Lie $so(2n)$ est d\'efinie par:

\begin{eqnarray*}
\ \ so(2n)=\left\{\left(
\begin{array}{cc}
X_{n\times n} & Y_{n\times  n} \\
Z_{n\times  n} & -X_{n\times  n}%
\end{array}%
\right), \quad \;\;  avec   \;\;\ {Y_{n\times n}+Y_{n\times
n}^{T}=0}, \quad {Z_{n\times  n}+Z_{n\times  n}^{T}=0}\right\}.
 \end{eqnarray*}
\item Nous avons deux propri\'et\'es pour $so(2n)$:
\begin{itemize}
\item  $so(2n)$ est une  sous al\`egbre de Lie de  $sl(2n,R )$
\item   La dimension de  $so(2n)$ est
$\frac{2n(2n-1)}{2}=n(2n-1)$.
\end{itemize}
\end{itemize}
\subparagraph{Exercices:}
\begin{enumerate}
\item Donner les relations de commutation de $so(4)$
\item  Montrer que $so(4)$ peut \'{e}tre vue comme deux copies de
$so(3).$
\end{enumerate}
Prenons maintenant  $J$ comme suit
 \begin{eqnarray*}
 J=\left(
\begin{array}{ccc}
1 & 0& 0 \\
0&0_{n\times n} & I_{n\times n} \\
0& I_{n\times n} & 0_{n\times n}
\end{array}%
\right)_{2n+1\times 2n+1}
 \end{eqnarray*}
Dans ce cas, on peut d\'{e}finir  $so(2n+1)$ comme suit
 \begin{eqnarray*}
 so(2n+1)=\left\{ X\in sl(2n+1,R),\quad JX+X^{T}J=0_{2n+1\times 2n+1}\right\}.
 \end{eqnarray*}
On consid\`ere  l'alg\`{e}bre  de Lie $so(3).$ Rappelons la forme
g\'{e}n\'{e}rale d'un \'{e}l\'{e}ment de $gl(3,{R})$ est comme suit
 \begin{eqnarray*}
  X=\left(
\begin{array}{ccc}
a & b & c \\
d & x & y \\
e & z & t%
\end{array}%
\right)
 \end{eqnarray*}
et prenons  la matrice  $J$  comme
 \begin {eqnarray*}
 J
=\left(
\begin{array}{ccc}
1 & 0 & 0 \\
0 & 0 & 1 \\
0 & 1 & 0%
\end{array}%
\right).
 \end{eqnarray*}
Comme auparavant, il faut r\'esoudre l'\'equation
$JX+X^{T}J=0_{3\times 3} $. En effet, nous  calculons l'expression
\begin{eqnarray*}
JX+X^{T}J &=&\left(
\begin{array}{ccc}
1 & 0 & 0 \\
0 & 0 & 1 \\
0 & 1 & 0%
\end{array}%
\right) \left(
\begin{array}{ccc}
a & b & c \\
d & x & y \\
e & z & t%
\end{array}%
\right) +\left(
\begin{array}{ccc}
a & d & e \\
b & x & z \\
c & y & t%
\end{array}%
\right) \left(
\begin{array}{ccc}
1 & 0 & 0 \\
0 & 0 & 1 \\
0 & 1 & 0%
\end{array}%
\right)  \\
&=&\left(
\begin{array}{ccc}
a & b & c \\
e & z & t \\
d & x & y%
\end{array}%
\right) +\left(
\begin{array}{ccc}
a & e & d \\
b & z & x \\
c & t & y%
\end{array}%
\right)  \\
&=&\left(
\begin{array}{ccc}
0 & 0 & 0 \\
0 & 0 & 0 \\
0 & 0 & 0%
\end{array}%
\right).
\end{eqnarray*}
Cela nous permet de trouver
\begin{eqnarray*}
a=0, \quad  b+e=0, \quad  c+d=0\\
x+t=0, \quad y=0, \quad  z=0.
\end{eqnarray*}
De la m\^eme facon que pour $so(2)$,  on  obtient   la solution
g\'{e}n\'{e}rale  suivante
 \begin{eqnarray*}
 X=\left(
\begin{array}{ccc}
0 & b & c \\
-c & x & 0 \\
-b & 0 & -x%
\end{array}%
\right).
\end{eqnarray*}
 Dans la base canonique $\{E_{ij}\}$, $X$  s'\'ecrit comme
\begin{eqnarray*}
X &=&x(E_{22}-E_{33})+b(E_{12}-E_{31})+c(E_{13}-E_{21}) \\
&=&xt_{1}+bt_{2}+c t_{3}.
\end{eqnarray*}
Pour \'{e}crire les relations de commutations, on peut utiliser  les
commutateurs  $\left[ E_{ij},E_{kl}\right] $.  En effet, dans une base form\'{e}e \ par les vecteurs
 $\left\{ \mid i>,\;i=1,2,3\right\} $,  la matrice $E_{ij}$ est
 repr\'esent\'ee par
\begin{eqnarray*}
 E_{ij}=\mid i><j\mid.
\end{eqnarray*}
Puisque le produit $ E_{ij}E_{k\ell} =\mid i><j\mid k><\ell\mid
=\delta _{jk}E_{i\ell}$, on obtient les relations de commutation
suivantes
\begin{eqnarray*}
\ \left[ E_{ij},E_{k\ell}\right] =\delta _{jk}E_{i\ell}-\delta
_{i\ell}E_{kj}\; avec\; i,j,k,\ell=1,2,3.
\end{eqnarray*}

\subparagraph{Exercices:}
\begin{enumerate}
\item  Donner les relations de commutation de l'alg\`{e}bre  de Lie $so(3)$.
\item V\'erifier que l'alg\`{e}bre $so(3)$ est isomorphe \`a  $su(2)$.
\end{enumerate}
 Finalement, pr\'esentons une derni\`ere  alg\`ebre  qui est l'alg\`{e}bre des  matrices symplectiques.  Pour cela, consid\'erons
   une autre forme  de $J$ comme suit
\begin{eqnarray*}
J=\left(
\begin{array}{cc}
0_{n\times n} & I_{n\times n} \\
-I_{n\times n} & 0_{n\times n}%
\end{array}%
\right) _{2n\times 2n}.
\end{eqnarray*}
L'alg\`{e}bre $sp(n, {C} )$ est d\'{e}finie par
\begin{eqnarray*}
g_{J}=\left\{ \left(
\begin{array}{cc}
X_{n\times n} & Y_{n\times n} \\
Z_{n\times n} & -X_{n\times n}^{T}%
\end{array}
\right), \qquad  Y=Y^{T}, \quad Z=Z^{T}\right\}.
\end{eqnarray*}
Notons que $sp(n)$  est une sous alg\`{e}bre de  $sl(2n,C)$.\\ Si on
consid\`ere l'alg\`ebre $sp(1,C)$,  on a
\begin{eqnarray*}
\ \ sp(1)=\left\{ \left(
\begin{array}{cc}
x & y \\
z & -x%
\end{array}%
\right) \right\}.
\end{eqnarray*}
{\bf Remarques}:\\
\begin{enumerate}
\item
Les g\'{e}n\'{e}rateurs de $sp(1)$ sont  $E_{11}-E_{22}$, $E_{21}$
et $E_{12}$. \item Sa   dimension est $\dim $
$sp(1)=3$\footnote{$sp(1)$ est isomorphe \`{a} $su(2)$}.
\item  Pour  $sp(n)$,  la dimension est
$n(2n+1)$.
\end{enumerate}

\subsection{ Classification des alg\`{e}bres de Lie}
Finalement,  nous donnons  une   classification pour les alg\`ebres
de Lie:  classification $ABCDEFG$. Elles sont class\'ees comme suit:
\begin{enumerate}
\item  Alg\`ebre de Lie   $A_n$: isomorphe \`a  $sl(n + 1,C)$,  ( $n\geq 1$)
\item  Alg\`ebre de Lie $B_n$: isomorphe \`a  $so(2n + 1)$,   ($n\geq 2$)
\item  Alg\`ebre de Lie $C_n$:  isomorphe  \`a  $sp(2n)$,  ($n\geq 3$)
\item  Alg\`ebre de Lie $D_n$:  isomorphe  \` a  $so(2n)$, ($n\geq 4$)
\item  Alg\`ebres  de Lie exceptionnelles:  $E_6$, $E_7$, $E_8$, $F_4$ et $G_2$.
\end{enumerate}

\chapter{ Alg\`{e}bre de Lie $sl(2,{C})$ et ses repr\'{e}sentations }
\pagestyle{myheadings}
\markboth{\underline{\centerline{\textit{\small{  Alg\`{e}bre de
Lie $A_{1}$ et ses
repr\'{e}sentations}}}}}{\underline{\centerline{\textit{\small{
Alg\`{e}bres de Lie $sl(2,{C})$ et ses repr\'{e}sentations}}}}}

Dans ce chapitre,  nous allons \'etudier l'alg\`{e}be  de Lie $sl(2,{C})$. En particulier,  nous rapplons la d\'efinition  de $sl(2,{C})$. En suite, nous donnons queleques repr\'esentations.
\section{Alg\`ebre de Lie  $sl(2,{C}$) }
L'alg\`{e}be  $sl(2,{C}) \simeq A_{1} $ est  une   alg\`ebre de Lie
de dimension trois   d\'{e}finie par les relations  de commutation
suivantes
\begin{eqnarray*}
\left[ h,e\right] &=&2e \\
\left[ h,f\right] &=&-2f \\
\left[ e,f\right] &=&h.
\end{eqnarray*}
{\bf Remarques}:
\begin{enumerate}
\item
Les g\'en\'erateurs  $\left\{ e,f,h\right\} $ forment   une base dite  la base de Cartan. \item Dans cette base, $%
h $ est dit un op\'erateur  qui compte les charges port\'{e}es par $e$ et $f$.\\
\item D\'eterminer une repr\'esentation   de  $ sl(2,C) $  revient \`a
r\'esoudre  sa structure abstraite.
\end{enumerate}
En fait, $sl(2,{C})$ poss\`ede
une infinit\'{e} de repr\'{e}sentations. Nous citons
\begin{itemize}
\item  Repr\'{e}sentations  matricielles
\item R\'{e}alisations   diff\'{e}rentielles.
\end{itemize}
\section{Repr\'{e}sentations de l'alg\`{e}bre de Lie $ sl(2,C) $ }
\subsection{R\'{e}alisations \ diff\'{e}rentielles}
Dans cette r\'{e}alisation, les g\'en\'erateurs  $h,f$ et $e$ sont repr\'{e}sent\'{e}s par des op\'{e}rateurs diff\'{e}rentielles agissant sur un
espace vectoriel dit espace de repr\'{e}sentation. En effet,  on
consid\`ere    l'espace vectoriel  $V=C \left[ X,Y\right] $ des
polyn\^omes en deux variables $X,Y$.  Pour $\lambda \geq 0$, on peut
d\'{e}finir  un espace vectoriel des polyn\^omes homog\'{e}nes de
degr\'{e} $\lambda $ $\left( V_{\lambda }\right):$
\begin{equation}
P_{\lambda }\left( X,Y\right) =\left\{ X^{\lambda },X^{\lambda
-1}Y,\ldots,Y^{\lambda }\right\} =\sum_{p+q=\lambda}
a_{p,q}X^{p}Y^{q}.
\end{equation}
Soit $\Phi $ un  homomorphisme   de l'alg\`ebre $ sl(2,C) $
\begin{equation}
\Phi :sl(2,C) \rightarrow gl(V_{\lambda }).
\end{equation}
A  travers $\Phi $, on trouve une    r\'ealisation  de
 $ sl(2,C) $ qui  est donn\'{e}e par la soultion suivante
\begin{eqnarray*}
&&e\rightarrow X\frac{\partial }{\partial
Y}=\Phi(e) \\
&&f\rightarrow Y\frac{\partial }{%
\partial X}=\Phi(f) \\
&&h\rightarrow X\frac{\partial }{\partial Y}%
-Y\frac{\partial }{\partial X}=\Phi(h).
\end{eqnarray*}
En effet, si nous calculons  le commutateur  $\left[ \Phi (e),\Phi
(f)\right]$, alors  on trouve que
$\left[ \Phi (e),\Phi (f)\right] =\Phi (e)\Phi (f)-\Phi (f)\Phi (e)=X\frac{%
\partial }{\partial X}$\ -$Y\frac{\partial }{\partial Y}=\Phi (h).$\\
{\bf Remarque}:\\
On peut donner aussi une autre r\'ealisation   sur  le plan des
nombres complexes $ C$  param\'etris\'e par  $ \left\{
z,\bar{z}\right\}$.   Cette solution est donn\'ee par les
op\'erateurs
\begin{eqnarray*}
&&e=z\frac{\partial }{\partial
\bar{z}} \\
&&f=\bar{z}\frac{%
\partial }{\partial z} \\
&&h=z\frac{\partial }{\partial \bar{z}} -\bar{z}\frac{\partial
}{\partial z}.
\end{eqnarray*}

\subsection{Repr\'{e}sentations matricielles}
Dans ce qui suit, on s'int\'eresse particuli\`erement aux repr\'esentations matricielles de $ sl(2,C) $.
La  repr\'{e}sentation   par des matrices
$ 2 \times $2  agissant sur un espace vectoriel de dimmension 2 est
appel\'{e}e repr\'esentation fondamentale. Cette derni\`ere est
donn\'ee par les matrices suivantes
\begin{equation}
e=\left(
\begin{array}{cc}
0 & 1 \\
0 & 0
\end{array}%
\right), \quad  f=\left(
\begin{array}{cc}
0 & 0 \\
1 & 0
\end{array}%
\right), \quad  h=\left(
\begin{array}{cc}
1 & 0 \\
0 & -1%
\end{array}%
\right).
\end{equation}
Notons qu'on peut  donner une   repr\'{e}sentation en  introduisant
les matrices de Pauli
\begin{equation}
\sigma _{1}=\left(
\begin{array}{cc}
0 & 1 \\
1 & 0%
\end{array}%
\right), \qquad  \sigma _{2}=\left(
\begin{array}{cc}
0 & -i \\
i & 0%
\end{array}%
\right),\qquad d   \sigma _{3}=\left(
\begin{array}{cc}
1 & 0 \\
0 & -1%
\end{array}%
\right)
\end{equation}
qui   v\'erifient  les  relations  suivantes
\begin{equation}
\left[ \sigma ^{i},\sigma ^{j}\right]
=i\sum_k\varepsilon^{ijk}\sigma ^{k},\qquad i,j,k=1,2,3
\end{equation}
avec  $\varepsilon^{ijk}$ est  le tenseur  compl\`etemnt
antisy\'etrique, et normalis\'e  de sorte que $\epsilon^{123}= +1$.
Le passage est assur\'e par  les relations  suivantes
\begin{eqnarray*}
h&=&\sigma _{3},\\   e&=& \frac{1}{2}(\sigma _{1}+i\sigma _{2})\\
 f&=&\frac{1}{2}(\sigma _{1}-i\sigma _{2}).
\end{eqnarray*}

\section{Les modules de $ sl(2,C) $}
Soit $g$  une alg\`{e}bre de Lie.  $V$ un espace
vectoriel muni d'une operation $\left( .\right) $ tel que:
\begin{eqnarray*}
g.V &\rightarrow &V \\
(x.v) &\rightarrow &x.v
\end{eqnarray*}
est dit $g$-module si nous avons les condutions
\begin{eqnarray*}
\left( ax+by\right) .v &=&ax.v+by.v \\
\left[ x,y\right] .v &=&xy.v-yx.v \\
x.(av+bw) &=&ax.v+bx.w,
\end{eqnarray*}
o\`u $a$ et $b$ sont des complexes.\\
{\bf Remarques:}
\begin{itemize}
\item On a  une compatibilit\'{e} entre $g$  et $V$
\item Le vecteur $x.v$ est d\'efini \`a travers un
homomorphisme
\begin{eqnarray*}
x.v=\Phi (x).v
\end{eqnarray*}
\item Le module adjoint est d\'{e}fini par $V=g$. Dans ce cas, on peut avoir
\begin{eqnarray}
g.g &\to &g \\
(x,y) &\to &x.y=xy-yx=\left[ x,y\right].
\end{eqnarray}
\end{itemize}
\subsection{D\'{e}finitions}
Un homomorphisme $\Phi$  assure que  les g\'en\'erateurs  de
l'alg\`{e}bre $ sl(2,C) $ sont
 des matrices $n\times n$ agissant $ sl(2,C) $ sur un espace vectoriel $V$ de dimension $n$. Pour chaque
 \'el\'ement $x$  de l'alg\`ebre  et un vecteur $v$  de $V$,  l'action   de $x$ sur  $v$ est donn\'ee par
\begin{eqnarray*}
x.v \rightarrow \Phi (x).v,
\end{eqnarray*}
o\`u $\Phi (x)$ est une matrice carr\'ee $n\times n$.\\
{\bf D\'{e}finition 1}: Les modules de $ sl(2,C) $ peuvent \^etre
d\'{e}compos\'es en somme directe des sous espaces propres de
dimension 1:
\begin{equation}
V=\oplus_{i}V_{i}.
\end{equation}
{\bf D\'{e}finition 2}: Un sous espace vectoriel $V_{\lambda _{i}}$
est d\'efini par $V_{\lambda _{i}}=\left\{ v_{i}\in V\ /\ \Phi
_{v}(h)v_{i}=\lambda_{i}v_{i}\right\} $ o\`u  $\lambda _{i}$ sont
les poids
de $h$ et $V_{\lambda_i }$.  On peut trouver les $\lambda _{i}$ en utlisant   l'action de $h$ sur les espaces des poides.
\\
{\bf Lemme}: Si $v\in V_{\lambda }$ alors on  a
\begin{itemize}
\item $e.v \in  V_{\lambda +2}$
\item $f.v \in V_{\lambda -2}$.
\end{itemize}
Autrement, nous avons aussi
\begin{eqnarray*}
e&:& V_{\lambda}\to V_{\lambda +2}. \\
f&:& V_{\lambda} \to V_{\lambda -2}.
\end{eqnarray*}

\subsection{Poid maximal et  vecteur maximal}
Dans la section suivante,  nous nous  int\'eressons  particuli\`erement aux représentations engendr\'ees par un vecteur ayant un  poid maximal. En particulier,
si $V$ est un $g$-module de $ sl(2,C) $ de dimension finie, et
$V=\oplus _{i}V_{\lambda _{i}}$ sa d\'ecomposition en sous espaces
de poids, alors il existe un poid maximal  $\lambda _{\max }$ tel
que
\begin{equation}
V_{\lambda _{\max }}\neq \left\{ 0\right\} \ \ \ \ \ \ \ \
V_{\lambda _{\max +2}}=\left\{ 0\right\}.
\end{equation}
Un vecteur $v_{0}$ est dit maximal si et  seulement si
\begin{eqnarray*}
hv_{0} &=&\lambda _{\max }v_{0} \\
e.v_{0} &=&0.
\end{eqnarray*}

\section{Modules irr\'{e}ductibles de $ sl(2,C) $}
Dans cette section, nous donnons une classification des modules irr\'{e}ductibles de $ sl(2,C) $. En effect,
on peut construire des sous modules \`a partir d'un vecteur maximal. Soit $v_{0}$ un vecteur maximal et $\lambda _{0}$ un poid
maximal tel que
\begin{eqnarray*}
hv_{0} &=&\lambda _{0}v_{0} \\
e.v_{0} &=&0.
\end{eqnarray*}
Posons
\begin{eqnarray*}
v_{n} &=&\frac{1}{n!}f^{n}v_{0} \\
v_{-1} &=&0,
\end{eqnarray*}
on obtient   le lemme suivant
\begin{eqnarray*}
h.v_{n} &=&(\lambda _{0}-2n)v_{n} \\
f.v_{n} &=&(n+1)v_{n+1} \\
e.v_{n} &=&(\lambda _{0}-n+1)v_{n+1}.
\end{eqnarray*}
{\bf D\'emonstration}: Exercice.
Pour chaque $A_{1}-$module irr\'{e}ductible de demension finie, nous
avons les propri\'{e}t\'{e}s suivantes:
\begin{itemize}
\item  Le param\'{e}tres $\lambda _{0}$ (poid maximal) est un entier
naturel. Autrement dit les modules de $ sl(2,C) $ sont class\'es par
un entier $m$. En effet, nous avons les situations suivantes
\begin{itemize}
\item  $m=0$: module scalaire

\item  $m=1$: module fondamentale

\item  $m=2$: module  adjoint.
\end{itemize}
\item Une base de $V_{m}$ est donn\'{e}e  par les vecteurs
\begin{equation}
\left\{ v_{n},\quad  n=0,\ldots m\right\}.
\end{equation}
\item La  dimension  de $V_{m}$ est $m+1$.
\end{itemize}
Consid\'erons  un  $g-$module de $ sl(2,C) $ de dimension $n+1$
\begin{equation}
V=C v_{0}\oplus C v_{1}\oplus\ldots \oplus C v_{n}
\end{equation}
tels que $v_{-1}=0$ et  $v_{n+1}=0$. Soit un vecteur $v_{n}$
d\'efini par
\begin{equation}
v_{n}=C \mid 0\rangle +C \mid 1\rangle +...C \mid n\rangle.
\end{equation}
L'action de $ sl(2,C) $ sur la base $\{\mid i\rangle, i=1,\ldots,n\}$ est  d\'efinie comme
suit
\begin{eqnarray*}
h \mid i\rangle &=&(n-2i)\mid i\rangle \\
f \mid i\rangle &=&(i+1)\mid i\rangle \\
e \mid i\rangle &=&(n-i+1)\mid i\rangle.
\end{eqnarray*}
{\bf Exemples}\\
Nous donnons  quelques exemples. Nous  commen\c cons  avec la
repr\'{e}sentation triviale  qui correspond \`a  $m=0$. Dons cas
cas,  la dimension de  $V$ est 1. Il est facil de voir que  l'action
de $ sl(2,C) $ sur $V$ est
\begin{eqnarray*}
h \mid 0\rangle &=&0\mid 0\rangle \\
f \mid 0\rangle &=&0\mid 0\rangle \\
e \mid 0\rangle &=&0\mid 0\rangle
\end{eqnarray*}
ceci conduit  ais\'ement \`a   la solution suivante
\begin{eqnarray*}
h  &=&0 \\
f &=&0 \\
e &=&0.
\end{eqnarray*}
Pour $m=1$,  la dimension de $V$    est 2:
\begin{eqnarray*}
V= C\mid 0\rangle \oplus C \mid 1\rangle.
\end{eqnarray*}
 Pour donner  la r\'ealisation  matricielle, Il  faut calculer  l'action  des  g\'en\'erateurs $e,f$ et $h$ sur
$V$. En effect, l'action de $e$ sur $V$ est donn\'ee par
\begin{eqnarray*}
e \mid 0\rangle &=&0\mid 0\rangle +0\mid 1\rangle \\
e \mid 1\rangle &=&1\mid 0\rangle +0\mid 1\rangle.
\end{eqnarray*}
Donc, le g\'en\'ereteur  $e$ est  repr\'esent\'e par la matrice
suivante
\begin{eqnarray*}
e=\left(
\begin{array}{cc}
0 & 0 \\
1 & 0%
\end{array}%
\right). \end{eqnarray*} Si nous calaculons  l'action de $f$, nous
obtenons
\begin{eqnarray*}
f \mid 0\rangle &=&0\mid 0\rangle +1\mid 1\rangle \\
f \mid 0\rangle &=&0\mid 0\rangle +0\mid 1\rangle
\end{eqnarray*}
et par cons\'equent on trouve
\begin{eqnarray*}f=\left(
\begin{array}{cc}
0 & 1 \\
0 & 0%
\end{array}%
\right). \end{eqnarray*} L'action de $h$ est comme suit
\begin{eqnarray*}
h \mid 0\rangle &=&1\mid 0\rangle +0\mid 1\rangle \\
h \mid 1\rangle & =&0\mid 0\rangle -1\mid 1\rangle,
\end{eqnarray*}
ceci nous donne
\begin{eqnarray*}h=\left(
\begin{array}{cc}
1 & 0 \\
0 & -1\end{array} \right). \end{eqnarray*} {\bf Exercice}: \'Etudier
le cas $m=2$.

\chapter{   Sous alg\`{e}bres de Cartan et syst\`{e}mes des racines}
\pagestyle{myheadings}
\markboth{\underline{\centerline{\textit{\small{ Sous alg\`{e}bres
de Cartan et syst\`{e}mes des
racines}}}}}{\underline{\centerline{\textit{\small{ Sous
alg\`{e}bres de Cartan et syst\`{e}mes des racines}}}}}
\section{Sous alg\`{e}bre de Cartan }
\subsection{Sous alg\`{e}bre Torique }
 On consid\`ere une alg\`{e}bre de Lie $g$ de dimension  finie.
On peut diviser les g\'{e}n\'{e}rateurs de $g$
en trois ensembles
\begin{itemize}
\item  Op\'{e}rateurs diagonaux $h_{i}$
\item Op\'{e}rateurs de cr\'{e}ation
 \item Op\'{e}rateurs  d'annihilation.
\end{itemize}
Soit $T$ un sous espace vectoriel
de $g$ engendr\'{e} par les op\'{e}rateurs diagonaux. \\
{\bf D\'efinition 1}:  $T$ est dite une alg\`{e}bre torique de
$g$.\\
{\bf D\'{e}finition 2}:
 Sous alg\`{e}bre de Cartan $H$ est la sous alg\`{e}bre torique
maximale de $g$.\\ {\bf Remarques}:
\begin{itemize}
\item
$H$ est une sous alg\`{e}bre de Lie  de $g$  ab\'elienne maximale
\begin{eqnarray*}
\left[ h_{i},h_{j}\right] =0.
\end{eqnarray*}
\item
La dimension de $H$ ($\dim H$)  est dite aussi  le rang de $g$
\begin{eqnarray}
\dim H=  rang \;  g.
\end{eqnarray}
\end{itemize}
{\bf Exemple 1: $sl(2,C )$}\\
Prenons l'alg\`ebre de Lie  $sl(2,C)$. Son alg\`ebre torique est
 donn\'ee par
\begin{equation}
T=\allowbreak \left\{ x_{s}=a\left(
\begin{tabular}{ll}
1 & 0 \\
0 & -1%
\end{tabular}%
\right),\;\;\;a\in C \right\}.
\end{equation}
$T$    est engendr\'{e}e par
\begin{equation}
h=\left(
\begin{tabular}{ll}
1 & 0 \\
0 & -1%
\end{tabular}%
\right).
\end{equation}
 Sa dimension   est   1.\\
{\bf Exemple 2: $sl(3,C )$}\\
  Rappelons que  la   dimension  $sl(3,C )$ est $3^{2}-1=8$.   Sa sous  alg\`{e}bre de
Cartan $H$  est donn\'ee par   la forme suivante
\begin{eqnarray*}
H=C h_{1}\oplus C h_{2},
\end{eqnarray*}
qui est aussi  la sous alg\`{e}bre  torique maximale de $sl(3,C )$. La
dimension de $H$ est 2
\begin{eqnarray*}
\forall h\in H, \;\;  h=\sum_{i=1}^{2} \alpha _{i}h_{i}, \quad \alpha _{i} \in C
\end{eqnarray*}
o\`u  $h_{1}=E_{11}-E_{22}$ et $h_{2}=E_{22}-E_{33}$.\\
{\bf Exemple 3: $sl(n+1,C )$}\\
 Dans ce cas, nous avons $n$ g\'en\'erateurs  diagonaux de type $h_i$. En termes des matrices $E_{ij}$, ils s'ecrivent  comme suit
\begin{eqnarray*}
h_{i}=E_{ii}-E_{i+1,i+1}, \qquad  i=1,\ldots, n.
\end{eqnarray*}
Dans ce cas, la dimension de $H$ est $n$:
\begin{eqnarray*}
H=C h_{1}\oplus C h_{2}\oplus\ldots\oplus C h_{n}.
\end{eqnarray*}
{\bf Exemple 4:} Pour  $sl(n,C)$, on a
\begin{eqnarray*}
\dim sl(n,C) &=&n^{2}-1 \\
rang \; sl(n,C)  &=&n-1
\end{eqnarray*}
{\bf Exemple 5:}  Pour $so(2n,R)$, on obtient
\begin{eqnarray*}
\dim so(2n) &=&\frac{2n(2n-1)}{2}=n(2n-1)\\
rang \;  so(2n)  &=&n.
\end{eqnarray*}

\section{D\'{e}composition de  Cartan}
Soit $g$ une alg\`{e}bre de Lie, on peut \'ecrire les relations suivantes
\begin{eqnarray*}
\forall x &\in &g , \quad  x=\sum \alpha _{a}x^{a} \\
\forall h &\in &H, \quad h=\sum \alpha _{i}h_{i} \\
ad_{h_{i}}(x_{a}) &=&\left[ h_{i},x_{a}\right] =\alpha
_{a}(h_{i})x_{a},
\end{eqnarray*}
o\`u $\alpha_a (h_i)$ sont les valeurs propres de $ad_{h_i}$.\\
\\
{\bf Exemple}: $sl(2,C )$. \\ Dans le cas de $sl(2,C )$,  $ad_{h}$
est une matrice $3\times 3$. En effet, dans la base de Cartan on a
\begin{eqnarray*}
ad_{h}(e) &=&\left[ h,e\right] =2e \\
ad_{h}(f) &=&\left[ h,f\right] =-2f \\
ad_{h}(h) &=&\left[ h,h\right] =0, \\
\end{eqnarray*}
et on obtient la matrice   $ad_{h}$
\begin{eqnarray*}
ad_{h} &=&\left(
\begin{tabular}{lll}
2 & 0 & 0 \\
0 & --2 & 0 \\
0 & 0 & 0%
\end{tabular}%
\right).
\end{eqnarray*}
{\bf Proposition}:La sous alg\`{e}bre de Cartan $H$ permet de
d\'{e}composer l'alg\`{e}bre de Lie en somme directe des espaces
propres.\\
\\
{\bf Lemme:} $g$  est d\'{e}composable en somme directe des sous
espace $g_{\alpha }$:
\begin{eqnarray*}
g &=&\oplus _{\alpha }g_{\alpha }
\end{eqnarray*}
o\`u
\begin{eqnarray*}
g_{\alpha } &=&\left\{ x\in g, \;\;\left[ h,x\right] =\alpha(h)x,
\;\;\forall h\in H\right\}.
\end{eqnarray*}
{\bf D\'efinition}: La d\'{e}composition $g=\oplus _{\alpha }g_{\alpha }$ est dite la d\'{e}%
composition de Cartan o\`u  $\alpha \geq 0$. \\
Autrement, nous \'ecrivons
\begin{eqnarray*}
g &=&g_{0}\oplus \left\{ \oplus _{\alpha \neq 0}g_{\alpha }\right\}
\\
g &=&H\oplus \left\{ \oplus _{\alpha \neq 0}g_{\alpha }\right\}.
\end{eqnarray*}
{\bf Exemple}:  $sl(2,C)$.\\
Rappelons que $sl(2,C)$ est engendr\'{e}e par $e,f$ et $h$ et on a
aussi
\begin{eqnarray*}
\dim A_{1} &=&3 \\
rang\; A_{1} &=&1.
\end{eqnarray*}
 La d\'{e}composition de Cartan de $sl(2,C)$ est donn\'ee par
\begin{eqnarray*}
sl(2,C) &=&   C h \oplus C e\oplus C
f \\
&=&g_{0}\oplus g_{+2}\oplus g_{-2},
\end{eqnarray*}
o\`u  $\ g_{0}=Ch$,  $ g_{-2}=Cf$  et  $g_{+2}=Ce$.\\
{\bf Remarque:} Dans la d\'{e}composition de Cartan, $\alpha $ est dite une forme
d\'{e}finie par $\alpha: H\to C$: c'est une application de sous
alg\`{e}bre de Cartan vers les valeurs complexes: $h\longrightarrow
\alpha (h)$.
\section{Forme de  Killing}
 La forme de Killing d'une
alg\`{e}bre de Lie $g$ est d\'{e}finie par
\begin{eqnarray*}
K:&g\times
g&\to C\\
&(x,y)&\to K(x,y)=Tr\; ad_{x} ad_{y}.
\end{eqnarray*}
La forme  $K(x,y)$ est:
\begin{itemize}
\item bilin\'{e}aire: $K(\alpha x+\beta
y,z)=\alpha
K(x,z)+\beta K(y,z),\;\;\; \alpha,\beta \in C$
\item sym\'{e}trique:  $K(x,y)=K(y,x)$.
\item associative dans le sens:  $ K(\left[ x,y\right] ,z)=K(x,\left[
y,z\right] )$.
\end{itemize}
{ \bf D\'{e}finition:} Soit $g$ une alg\`{e}bre de
Lie.  $g$ est dite semi simple si la forme de Killing $K(x,y)$ est
non d\'{e}g\'{e}ner\'{e}e ($\det $ $K\neq 0$) o\`u $K$ est la
matrice associ\'{e}e \`{a} la forme de Killing.\\
\\{\bf Exemple:}
 $g=sl(2,C)$.\\
 Rappelons que  les relations de commutation de
 $g=sl(2,C)$  sont
\begin{eqnarray*}
\left[ h,e\right] &=&2e \\
\left[ h,f\right] &=&-2f \\
\left[ e,f\right] &=&h.
\end{eqnarray*}
 Dans ce cas, la matrice  $K$ associ\'{e}e  \`{a} la forme de Killing est donn\'ee
par
\begin{eqnarray*}
K=\left(
\begin{array}{ccc}
K(e,e) & K(e,f) & K(e,h) \\
K(f,e) & K(f,f) & K(f,h) \\
K(h,e) & K(h,f) & K(h,h)%
\end{array}%
\right).
\end{eqnarray*}
En effet,  on   a $ ad_{h}(e)=2e+0f+0h$,  $ad_{h}(f)=0e-2f+0h$ et
$ad_{h}(h)=0e+0f+0h$. Par cons\'equent, on obtient les matrices $ad_{h,e,f}$
\begin{eqnarray*}
ad_{h}=\left(
\begin{array}{ccc}
2 & 0 & 0 \\
0 & -2 & 0 \\
0 & 0 & 0%
\end{array}%
\right), \quad  ad_{f}=\left(
\begin{array}{ccc}
0 & 0 & -1 \\
0 & 0 & 0 \\
2 & 0 & 0%
\end{array}%
\right),  \quad  ad_{e}=\left(
\begin{array}{ccc}
0 & 0 & 0 \\
0 & 0 & 1 \\
-2 & 0 & 0%
\end{array}%
\right). \end{eqnarray*}  Finalement, la matrice  $K$ associ\'{e}e  \`{a} la forme
de Killing est donn\'ee par
 \begin{eqnarray*}
 K=\left(
\begin{array}{ccc}
0 & 2 & 0 \\
2 & -4 & 0 \\
0 & 0 & 8%
\end{array}%
\right).
\end{eqnarray*}
\section{Syst\`eme de racines}
 Soit $g$ une alg\`{e}bre de Lie de dimension finie et $H$ sa sous
alg\`{e}bre de Cartan.\\
{\bf D\'efinition:}  L'ensemble des formes $\alpha $ non nulles
$(g_{\alpha }\neq 0)$
\begin{eqnarray*}
\Delta =\left\{ \alpha \in H^{\ast }/g_{\alpha }\neq 0\right\}
\end{eqnarray*} est appel\'{e} syst\`{e}me des racines.\\
Consid\'{e}rons  maintenant la d\'{e}composition de Cartan
 \begin{eqnarray*}
g=H\oplus \left\{ \oplus _{\alpha \in \Delta }g_{\alpha }\right\}.
 \end{eqnarray*}
Nous avons les propri\'et\'es suivantes
\begin{itemize}
\item
$\left[ g_{\alpha },g_{\beta }\right] \subset g_{\alpha +\beta }$
\item si
$x\in g_{\alpha }(\alpha \neq 0)$,  alors $ad_{x}^{2}$ est nul
\item Si $\alpha $ et $\beta \in H^{\ast }$ et $\alpha +\beta \neq 0$, alors $g_{\alpha }$ $\perp g_{\beta }$
relativement \`{a} la forme de Killing.
 \end{itemize}
\subsection*{D\'{e}monstration:}
Si  $x\in g_{\alpha }$ et $y\in g_{\beta }$, alors $ad_{h}(x)=\alpha
(h)x$ et $ad_{h}(y)=\beta (h)y$.\\ Calculons  $ad_{h}(\left[
x,y\right] )$. En effet, on a
\begin{eqnarray*}
ad_{h}(\left[ x,y\right] ) &=&\left[ ad_{h}(x),y\right] +\left[ x,ad_{h}(y)%
\right] \\
&=&\left[ \alpha (h)x,y\right] +\left[ x,\beta (h)y\right] \\
&=&\alpha (h)\left[ x,y\right] +\beta (h)\left[ x,y\right] \\
&=&(\alpha +\beta )(h)\left[ x,y\right]
\end{eqnarray*}
 qui implique que $ \left[ x,y\right] \subset g_{\alpha +\beta }$.\\
Calculons $ad_{x}^{2}(h)$.  En effet, on trouve  que
\begin{eqnarray*}
ad_{x}^{2}(h) &=&ad_{x}(ad_{x}(h)) \\
&=&ad_{x}(\alpha (h)x) \\
&=&\alpha (h)ad_{x}(x) \\
&=&\alpha (h)\left[ x,x\right] \\
&=&0.
\end{eqnarray*}
D\'emonstration de la derni\`ere propri\'et\'e: \\ Consid\'erons un \'el\'ement
 $x$ de $ g_{\alpha }$ alors $ ad_{h}(x)=\alpha (h)x$
et si   $y\in g_{\beta }$ alors $ ad_{h}(y)=\beta (h)y$. Il  faut
d\'emontrer  que $K(x,y)=0$  si  $\alpha +\beta \neq 0$.  Pour cela,
on va utiliser la   propri\'et\'e suivante
\begin{eqnarray*}
K(h,\left[ x,y\right] )=K(\left[ x,y\right],h).
\end{eqnarray*}
En effet, nous calculons
\begin{eqnarray*}
\alpha (h)K(x,y) &=&K(\alpha (h)x,y) \\
&=&K(\left[ h,x\right] ,y) \\
&=&K(x,\left[ y,h\right] ) \\
&=&-\beta (h)K(x,y)
\end{eqnarray*}
 ceci nous donne  $ \left( \alpha (h)+\beta (h)\right) K(x,y)=0$. Si
$\alpha +\beta \neq 0$ alors $ K(x,y)=0$ et par cons\'equent
$ g_{\alpha }$ et $g_{\beta }$ sont perpendiculaires relativement \`a la forme de  Killing.\\
{\bf Propri\'{e}t\'{e}s de $\Delta $}.\\
Nous pr\'esentons  quelques propri\'et\'es de  $\Delta $:
\begin{enumerate}
\item  Orthogonalit\'{e}
\item Int\'{e}grabilit\'{e}
\item Rationalit\'{e}
\end{enumerate}
{\bf Orthogonalit\'{e}}: Nous donnons  les  propri\'et\'es suivantes
\begin{itemize}
\item  $\Delta $ engendr{e} $H^{\ast }.$
\item  Si $\alpha \in \Delta $,
$-\alpha \in \Delta.$
\item  Soit $\alpha \in
\Delta $.  Si  $x\in g_{\alpha }$ et  $y\in g_{-\alpha }$ \ alors
\begin{eqnarray*}
\left[ x,y\right] =K(x,y)t_{\alpha }, \qquad t_{\alpha }\in H
\end{eqnarray*}
et on  a  $K(t_{\alpha },h)=\alpha (h)$
 \item Si $\alpha \in \Delta$ alors $ \left[ g_{\alpha },g_{-\alpha }%
\right] $ est un sous espace vectoriel de dimension 1  engendr\'e
par  $t_{\alpha }$ tel que
\begin{eqnarray*}
\left[ g_{\alpha },g_{-\alpha }\right] \simeq Ct_{\alpha }.
\end{eqnarray*}
\item Si $\alpha \in \Delta $, $x_{\alpha }\in g_{\alpha
}$, il exite un \'el\'ement  $y_{\alpha }\in g_{-\alpha }$ tel que $
\left\{ x_{\alpha },y_{\alpha },h_{\alpha }\right\} \simeq sl(2,C)$
et
$h_{\alpha }=\frac{2t_{\alpha }}{\alpha (t_{\alpha })}=\frac{2t_{\alpha }}{%
K(t_{\alpha },t_{\alpha })}$.
\end{itemize}
{\bf Exemple $sl(3,C)$}: \\
Revenons  \`a  $sl(3,C)$. On sait que sa
dimension de  $sl(3,C)$ est $8$ et  son $rang$ est $2$. Sa  sous
alg\`ebre de Cartan   $H$ est donn\'ee par
\begin{eqnarray*}
Ch_{1}\oplus Ch_2.
\end{eqnarray*}
On obtient $\left\vert \Delta \right\vert =\dim $ $sl(3,C)-rang$
$sl(3,C)=8-2=6$. Le syst\`eme des racines positives  est $\Delta
_{+}=\left\{ \alpha _{1},\alpha _{2},\alpha _{1}+\alpha _{2}\right\}
$. Puisque que l'opos\'e d'une racine est aussi une racine, les racines n\'egatives sont $\Delta _{-}=\left\{ -\alpha _{1},-\alpha
_{2},-(\alpha _{1}+\alpha _{2})\right\} $. La d\'ecomposition de
Cartan  de $sl(3,C)$ est donn\'ee
\begin{eqnarray*}
sl(3,C)=H\oplus g_{\pm \alpha _{1}}\oplus g_{\pm \alpha _{2}}\oplus
g_{\pm (\alpha _{1}+\alpha _{2})}.
\end{eqnarray*}
{\bf Rationalit\'{e}}:\\
Soit  $g$ une alg\`{e}bre de Lie,  $H$ sa sous alg\`{e}bre de Cartan
 et  $H^{\ast }$  est le  dual de $H$. Rappelons que
la d\'{e}composition de Cartan est
\begin{eqnarray*}
g=H\oplus \left\{ \oplus _{\alpha \in \Delta }g_{\alpha
}\right\},\quad g_{\alpha }=\left\{ x\in g/\left[ h,x\right] =\alpha
(h)x\right\}.
\end{eqnarray*}
{\bf Lemme:}\\
On consid\`ere une base  de $%
H^{\ast }$ form\'{e}e par des racines de $\Delta =\left\{   \left(
\alpha _{1},\ldots,\alpha _{\ell}\right)\right\} $. Alors,  on a les
deux propri\'et\'es suivantes
\begin{enumerate}
\item $\beta =\sum\limits_{j=1}^{\ell}C_{j}\alpha _{j}, \qquad
\forall \beta \in \Delta$
\item  $C_{j}\in  Q$.
\end{enumerate}

{\bf Integrabilit\'{e}}: Nous avons les  propri\'et\'es suivantes
\begin{enumerate}
\item Si $\alpha \in \Delta $, alors $\dim $ $g_{\alpha }=1$
\item Si $\alpha \in \Delta $,  $k\alpha \in \Delta $ si $k=\pm 1$
\item  Si $\alpha \in \Delta $,  $\beta \in \Delta $ et $\alpha +\beta  \in \Delta $,
 alors $[ g_\alpha,g_\beta]=g_{\beta}$
\item     Si $\alpha \in \Delta $ et   $\beta \in \Delta $, alors $K(t_\alpha,t_\beta)=\beta(h_\alpha)\in Z$.
\item  $g$ est engendr\'ee par $g_\alpha$.
\end{enumerate}
{\bf Exemple: $sl(2,C)$}. Dans ce  cas, on a
\begin{eqnarray*}
sl(2,C)&=&C h\oplus C e\oplus C
f \\
&=&C h\oplus C g_{\alpha }\oplus C g_{-\alpha }
\end{eqnarray*}
o\`u
\begin{eqnarray*}
g_{\alpha } =Ce, \qquad g_{-\alpha } =Cf.
\end{eqnarray*}.
{\bf Remarque:} Notons que $\dim g_{\alpha }=\dim g_{-\alpha }=1$.

\chapter{   Syst\`{e}me des racines}
\pagestyle{myheadings}
\markboth{\underline{\centerline{\textit{\small{ Syst\`{e}me des
racines}}}}}{\underline{\centerline{\textit{\small{ Syst\`{e}me des
racines}}}}}En th\'eorie des alg\`ebres de Lie, un syst\`eme des
racines est une configuration
   de vecteurs  dans un espace Euclidien v\'erifiant certaines conditions
      g\'eom\'etriques. Dans ce chapitre, nous discutons les points suivants:
\begin{enumerate}
\item  Espaces Euclideans.
\item  R\'eflexions.
\item Syst\`emes  des Racines.
\end{enumerate}
\section{Espace Euclidien}
{\bf D\'efinition 1}\\
Soit  $E$  un espace vectoriel de dimension finie sur le corps des nombres r\'eels  $R$.
 Une application $(.)$  d\'efinie positive
\begin{eqnarray*}
(,) : E\times E & \to & R
\\\alpha,\beta & \to & (\alpha , \beta )=\alpha.\beta
\end{eqnarray*}
est dite  produit scalaire s'elle est   \\ $\bullet $
bilin\'{e}aire
\begin{center}
$\left( a \alpha +b \gamma,\beta \right) =a \left( \alpha ,\beta
\right) +b \left( \gamma,\beta \right) $
\end{center}
$\bullet $ sym\'{e}trique
\begin{center}
$(\alpha,\beta )=(\beta ,\alpha )$
\end{center}
$\bullet $ d\'{e}fini positive
\begin{center}
$(\alpha, \alpha )$ $>$ $0$ \ \ et \ \ $(\alpha, \alpha )=0$ \ \ si \ \ $%
\alpha =0$
\end{center}
pour tout $\alpha, \beta \in E$ et $a,b \in R$.\\
{\bf Remarques:}
\begin{enumerate}
\item
 La notion  du  produit scalaire est valable en dehors du cadre de la dimension finie. Elle se g\'en\'eralise \'egalement
  aux espaces complexes.
\item Le produit scalaire permet de d\'efinir une norme et une distance.
\end{enumerate}
{\bf D\'efinition 2}
Un espace euclidien est un espace vectoriel r\'eel de dimension finie muni d'un produit scalaire.\\
{\bf Exemple}: Si $E=R^n$, il s'agit de la structure Euclidienne
naturelle sur $R^n$. Dans ce cas, le produit scalaire est dit   canonique
\begin{eqnarray*}
(\alpha,\beta)=\alpha.\beta=\alpha_1\beta_1+ \alpha_2\beta_2+
\ldots+\alpha_n\beta_n.
\end{eqnarray*}
Dans ce qui suit, $E$  est un espace euclidien de dimension $\ell$ ($\dim \ell$ $(\ell<\infty )$).
\subsection{R\'{e}flexions}
 On d\'efinit une r\'{e}flexion $\sigma $ dans $E$  comme une
application lin\'{e}aire $E\to E$  donne\'{e}e par
\begin{eqnarray*}
\sigma _{v}: &  E&\to  E\\
&w&\to \sigma _{v}(w)=w-\frac{2(w.v)v}{(v.v)},\quad  \forall v\in E
\end{eqnarray*}
o\`u  $w.v$ d\'esigne le produit scalaire dans $E$.\\ Cette
r\'eflexion poss\`ede  certaines
 propri\'et\'es:
\begin{enumerate}
\item  $\sigma^2 _{v}(w)=w$
\item  Si $w\perp v$, alors $\sigma
_{v}(w)=w$  et par  cons\'equent $ w$ est invariant par $\sigma _{v}$.
\item On a l'expression  suivante
\begin{center}
$\sigma _{v}(v)=v-\frac{2(v.v)v}{(v.v)}=-v,\;\;\;\;\;\sigma
_{v}(\lambda v)=-\lambda v$ \ $\;$ \ \ $\lambda \in R .$
\end{center}
\item La r\'{e}flexion conserve le produit scalaire c'est \`{a} dire
\begin{center}
$(\sigma _{v}(\alpha ).\sigma _{v}(\beta ))=(\alpha .\beta ).$
\end{center}
\end{enumerate}
\subsubsection{D\'{e}monstration }
Ici,  nous d\'emontrons seulement   la propri\'et\'e  3.  On commence par  calculer le produit scalaire  $\sigma _{v}(\alpha ).\sigma _{v}(\beta )$.
En effet, nous avons
\begin{eqnarray*}
\sigma _{v}(\alpha ).\sigma _{v}(\beta ) &=&(\alpha -\frac{2(\alpha .v)v}{%
(v.v)}.\beta -\frac{2(\beta .v)v}{(v.v)}) \\
&=&(\alpha .\beta )-\frac{2(\alpha .v)(\beta
.v)}{(v.v)}-\frac{2(\alpha
.v)(v.\beta )}{(v.v)}+\frac{4(\alpha .v)(\beta .v)}{(v.v)^{2}}(v.v) \\
&=&(\alpha .\beta ).
\end{eqnarray*}
Pour les autres propri\'et\'es, on peut se r\'ef\'erer au live [1].
\section{ Syst\`emes des Racines}
Soit $\Delta $ un
sous ensemble de $E$. On appelle  $\Delta $  un syst\`{e}me des
racines de $E$ si nous avons les conditions suivantes
\begin{itemize}
\item $\Delta $ est fini, engendre $E$, $0\notin \Delta $
\item $\forall \alpha \in \Delta $, $k\alpha \in \Delta $  si $%
k=\pm 1$
\item
 $\forall \alpha \in \Delta $, $\sigma _{v}$ stabilise
$\Delta$, autrement   $\sigma _{\alpha }(\Delta )\subset \Delta $
\item  Si $\alpha $ et $\beta $
dans $\Delta$, le produit scalaire
\begin{eqnarray*}
\left\langle \beta .\alpha \right\rangle &=&\frac{2(\beta .\alpha
)}{\left( \alpha .\alpha \right) }\in Z.
\end{eqnarray*}
\end{itemize}
{\bf Remarque}: Si $\left( \alpha, \alpha \right)=2$,  alors le produit
scalaire $\left\langle \beta .\alpha \right\rangle$ devient produit scalaire usuel
sym\'etrique. En effet,
\begin{eqnarray*}
\left\langle \beta .\alpha \right\rangle =\alpha.\beta =\beta
.\alpha.
\end{eqnarray*}
\subsubsection*{Exemple:}
Soit $E$ un espace vectoriel de dimension $1$ tel que
\begin{center}
$E=C \alpha,\;\;\;\;\;\;\Delta =\left\{ \alpha ,-\alpha \right\}. $
\end{center}
Rappelons que la dimension de E est:
\begin{center}
$\dim E=rang$ $g=\dim H$.
\end{center}
Le  nombre des \'{e}l\'{e}ments de $\Delta $ est:
\begin{center}
$\left\vert \Delta \right\vert =2$
\end{center}
qui montre que $\Delta $ est fini.   Il est  facile de voir que
\begin{center}
$\sigma _{\alpha }(\alpha )=-\alpha \in \Delta, \;\;\;\;\;\sigma
_{\alpha }(-\alpha )=+\alpha \in \Delta $
\end{center}
donc $\ \ \sigma _{\alpha }(\Delta )\subset \Delta $  et d'autre
part, on a aussi
\begin{center}
$\sigma _{-\alpha }(\alpha )=-\alpha,\;\;\; \sigma _{-\alpha
}(-\alpha )=+\alpha $
\end{center}
et par cons\'equent
\begin{center}
$\sigma _{-\alpha }(\Delta )\subset \Delta. $
\end{center}
Pour  tout $ \alpha \in \Delta$,  on a  $ \sigma
_{\alpha }(\Delta )\subset \Delta $,  et nous avons le r\'esultat attendu.\\La derni\`ere propri\'et\'e de
$\Delta$ est assur\'ee par
\begin{eqnarray*}
\left\langle \alpha ,\alpha \right\rangle &=&\frac{2 \alpha. \alpha  }{ \alpha.\alpha}=2\in Z
\\
\left\langle -\alpha,\alpha \right\rangle &=&-2\in Z
\\
\left\langle \pm \alpha ,\alpha \right\rangle &=&\pm 2\in Z.
\end{eqnarray*}
\section{Groupe de Weyl } Consid\'erons   un syst\`{e}me des racines
$\Delta $. L'ensemble des r\'eflexions dans $E$
\begin{eqnarray*}
W=\left\{ \sigma _{\alpha }:\Delta \longrightarrow \Delta  \;\;
\forall \alpha \in \Delta \right\}
\end{eqnarray*} a une structure de groupe et   dit  le groupe de Weyl associ\'{e} \`{a} $\Delta $.
\subsection*{Propri\'{e}t\'{e}s de $W$ }
\begin{itemize}
 \item $W$ est un sous groupe de $GL(E).$
\item $W$ est fini ($\Delta $ est fini).
\item $W$ est  un groupe  de permutation qui permute les racines (\'{e}l\'{e}ments de $\Delta$). On peut  dire aussi que $W$ est un   groupe de r\'efelxions associ\'e \`a $\Delta$.
\end{itemize}
\subsubsection*{Exemples}
Prenons comme exemple l'alg\`{e}bre $sl(2, C)$.  Sa  dimension  est
$\dim A_{1}=3$ et son rang est 1. Le nombre des \'{e}l\'{e}ments de
$\Delta$ est
\begin{center}
 $\left\vert \Delta \right\vert =\dim A_{1}-rangA_{1}=2$
\end{center}
En effet, nous avons
\begin{center}
$\Delta =\Delta _{+}\cup \Delta _{-}=\left\{ +\alpha ,-\alpha
\right\} $
\end{center} o\`u
$\Delta _{+}=\left\{ +\alpha \right\} $ et  $\Delta _{-}=\left\{
-\alpha \right\} $.\\  Puisque   $\left\vert \Delta \right\vert =2<\infty
$, alors  $\Delta $ est fini. Dans ce cas, on peut \'ecrire
$E$ comme
\begin{eqnarray*}
E\simeq C \alpha.
\end{eqnarray*}
On a \'egalement
\begin{center}
$\alpha ,-\alpha ;\sigma _{\pm \alpha }(\Delta )\subset \Delta $
$\left\langle \alpha ,\pm \alpha \right\rangle =\pm 2\in Z $
\end{center}
On peut voir facilement  que
\begin{center}
$\sigma _{\alpha }^{2}(\alpha )=\sigma _{\alpha }.\sigma _{\alpha
}(\alpha )=\sigma _{\alpha }(-\alpha )=\alpha, \qquad
\sigma_{-\alpha }^{2}(\alpha )=\alpha. $
\end{center}
$W$ a deux \'{e}l\'{e}ments. Il est donc comme le groupe cyclique $Z
_{2}$ o\`u $Z_{2}=\left\{ e,g\right\} $ tel que\footnote{Voir
chapitre 6}
\begin{center}
$g.g=e$
\end{center}
Pour l'alg\`ebre $sl(n+1, C)$, le groupe  de r\'eflexions   est
isomophe \`a $S_n$ (le groupe sym\'etrique \`a $n$ \'el\'ements).
\section{Classification  des racines}
\subsection{Positions relatives des racines}
Soient $\alpha $ et $\beta $ deux racines de  $(\Delta )$. Rappelons que le
produit scalaire   entre  $\alpha $ et $\beta $ est  alors  d\'efini
par
\begin{eqnarray*}
\left\langle \beta ,\alpha \right\rangle =\frac{2\beta.\alpha}{\alpha.\alpha}.
\end{eqnarray*}
{\bf Remarque}: Dans le cas o\`u  $\left\vert \alpha \right\vert
=\left\vert^2 \beta \right\vert^2$, on trouve que
\begin{center}
$ \left\langle \beta ,\alpha \right\rangle =(\alpha,\beta)$
\end{center}
Consid\'erons maintenant  l'angle $\theta $  entre $\alpha $ et
$\beta $ ($\theta =\hat{\left( \alpha ,\beta\right) }$)  pour tout
$\alpha $ et $\beta$ de $\Delta $. Effet,  nous calculons l'expression
\begin{eqnarray*}
\left\langle \beta ,\alpha \right\rangle &=&\frac{2(\beta ,\alpha )}{%
(\alpha ,\alpha )} \\
&=&\frac{2\left\vert \alpha \right\vert \left\vert \beta \right\vert
\cos
\theta }{\left\vert \alpha \right\vert ^{2}} \\
&=&\frac{2\left\vert \beta \right\vert \cos \theta }{\left\vert
\alpha \right\vert }
\end{eqnarray*}
et on obtient
\begin{eqnarray*}
\frac{\left\langle \beta ,\alpha \right\rangle}{\left\langle \alpha
,\beta \right\rangle}=\frac{\left\vert \alpha
\right\vert^2}{\left\vert \beta \right\vert^2} \geq 0.
\end{eqnarray*}
On  constate alors que $\left\langle \alpha ,\beta \right\rangle $
et $\left\langle \beta,\alpha \right\rangle $ ont le m\^{e}me signe.
De m\^eme, on a
\begin{eqnarray*}
\left\langle \beta ,\alpha \right\rangle \left\langle \alpha ,\beta
\right\rangle =4\cos ^{2}\theta
\end{eqnarray*}
qui conduit \`a
\begin{eqnarray*}
0 &\leq &\left\langle \beta ,\alpha \right\rangle \left\langle
\alpha ,\beta \right\rangle \leq 4.
\end{eqnarray*}
On a aussi  les quatres situations   suivantes
\begin{center}
$\left\langle \beta ,\alpha \right\rangle \left\langle \alpha ,\beta
\right\rangle =0,1,2,3,4.$
\end{center}
Il se trouve que les  valeurs  possibles de  $\theta$ sont
$30^{0},45^{0},60^{0},90^{0},120^{0},135^{0},150^{0}.$
\subsection{Syst\`{e}me de racines de $rang$ $1$}
Il existe seulement un syst\`{e}me de racines de $rang$ $1$ qui
contient  deux vecteurs (racines) $\alpha $ et $-\alpha $. Ce
syst\`{e}me de racine est appel\'{e} $A_{1}.$
\subsection{Syst\`{e}me de raciness de $rang$ $2$}
Pour le rang 2, il y a  quatre  possibilit\'es:\\
  $\bullet $ $so(4)\simeq su(2)\oplus su(2)$:\\
 Dans ce cas,  on a
\begin{center}
$\alpha \perp \beta, \qquad  \left\langle \alpha ,\beta
\right\rangle =\left\langle \beta ,\alpha \right\rangle =0$
\end{center}
et
\begin{center}
$\sigma _{\alpha }(\beta )=\beta, \;\;\;\; \sigma _{\beta }(\alpha
)=\alpha $
\end{center}
Pour cela, on trouve  que
\begin{center}
$\theta =\frac{\pi }{2}.$
\end{center}
$\bigskip $ $\bullet $ $sl(3,C)$:\\
Dans ce cas, nous avons
\begin{center}
$\left\langle \alpha ,\beta \right\rangle =\left\langle \beta
,\alpha \right\rangle =-1$
\end{center}
et donc
\begin{center}
$\theta =\frac{2\pi }{3}.$
\end{center}
Pour cela, v\'erifions que   $\alpha+\beta$ est aussi une racine. En
effet, on obtient
\begin{center}
$\sigma _{\alpha }(\beta )=\beta -\left\langle \beta ,\alpha
\right\rangle \alpha =\beta +\alpha $
\end{center}
alors le syst\`eme de racines $\Delta$  est
\begin{center}
$\Delta =\left\{ \pm \alpha ,\pm \beta ,\pm (\alpha +\beta )\right\}
$
\end{center}
$\bullet $ $so(5)$:\\
Pour cet alg\`ebre, nous avons
\begin{center}
$\left\langle \beta ,\alpha \right\rangle =-2$ \ et \ $\left\langle
\alpha ,\beta \right\rangle =-1$
\end{center}
d'o\`{u}
\begin{center}
$\Delta =\left\{ \pm \alpha ,\pm \beta ,\pm \left( \alpha +\beta
\right) ,\pm \left( \beta +2\alpha \right) \right\}. $
\end{center}
$\bullet $ $G_{2}$:\\
Dans cette situation,  on a
\begin{center}
$\left\langle \beta ,\alpha \right\rangle =-3$ \ et \ $\left\langle
\alpha ,\beta \right\rangle =-1$.
\end{center}
on sait que
\begin{center}
$\dim G_{2}=14$ \ \ et \ \ \ $rang\;G_{2}=2$
\end{center} et par cons\'equent
\begin{center}
$\left\vert \Delta \right\vert =14-2=12=2\times 6.$
\end{center}
Les \'el\'ements de $\Delta$ sont comme suit
\begin{center}
$\Delta =\left\{ \pm \alpha ,\pm \beta ,\pm (\alpha +\beta
), \pm (2\alpha +\beta
),\pm (3\alpha +\beta
), \pm (3\alpha +2\beta)     \right\}.$
\end{center}
\section{Racines simples} Soit $\Delta $ un syst\`{e}me des racines
d'un espace vectoriel $E$ et $W$ son groupe de Weyl tel que
\begin{center}
$rang\;\Delta =\dim E=\ell =rang$ $g=\dim H$
\end{center}
\subsubsection{Base de racines $\protect\pi $:}
Une partie $\pi $ tel que
\begin{center}
$\pi =\left\{ \alpha _{i}, \;\;i=1,\ldots \ell \right\} \subset
\Delta $
\end{center}
est dite une base de $\Delta $, si nous avons les conditions
suivantes:\begin{itemize}
\item $\pi $ est une base de $E$
\item $\forall \beta \in \Delta $ on a $\beta =\sum\limits_{i=1}^{\ell}$ $%
k_{i}\alpha _{i}$ \ ou \ $\left\{ k_{i}<0, k_{i}>0\right\} $, o\`u
$k_{i}$ sont des entiers de m\^{e}me signe.
\end{itemize}
{\bf Remarques:}
\begin{enumerate}
\item Si $k_{i}>0$,  $  \beta $ une racine positive  et si
$k_{i}<0$ alors $\beta $ une racine n\'{e}gative.
\item  Les \'{e}l\'{e}ments de $\pi $ sont appel\'{e}s racines simples
($\alpha _{i}=\sum k_{j}\alpha _{j}\notin \pi$).
\item Les racines simples permettent de d\'eterminer totalemnt l'alg\`ebre.
\end{enumerate}
{\bf Propri\'{e}t\'{e}s :}\\
$\bullet $ Au lieu de dire une base on dit un syst\`{e}me de racines simples $\pi.$\\
$\bullet $ $Card \pi =rang$ $g=\dim H=\dim E=\ell$.\\
$\bullet $ La d\'{e}composition \ $\beta =\sum\limits_{i}k_{i}\alpha
_{i}$ est unique.\\
\\
{\bf Exemple :} Nous illustrons  l'exemple  $A_{2}$. Les
 \'el\'ements de $\Delta$ (racines) sont
\begin{center}
$\Delta (A_{2})=\left\{ \pm \alpha ,\pm \beta ,\pm (\alpha +\beta
)\right\} $
\end{center}
On remarque que  les racines simples sont
\begin{center}
$\pi (A_{2})=\left\{ \alpha, \beta \right\}. $
\end{center}
Pour terminer ce chapitre, nous allons  pr\'esenter  des propositions sur $\pi$.  Cependant,
 pour plus de d\'etails, on peut se r\'ef\'erer aux livres [1,2].\\
{\bf Propositions}:\\
 $\bullet $ $\left\langle \alpha _{i},\alpha _{j}\right\rangle
\leq 0$ \\$\bullet $ $\alpha _{i}-\alpha _{j}\in \Delta $.\\
 $\bullet
$ $W$ est un groupe de Weyl engendr\'{e} par $\sigma _{\alpha_i}.$\\
$\bullet $ Soient $\pi $ et $\pi ^{\prime }$ deux bases de $\Delta $
et  $W$ est le groupe de Weyl associ\'{e} \`{a} $\Delta $, alors il
existe $\sigma \in W$ telle que $ \sigma(\alpha) \in \pi.$

\chapter{    Matrices de Cartan et diagrammes de Dynkin}
\pagestyle{myheadings}
\markboth{\underline{\centerline{\textit{\small{Matrices de Cartan
et diagrammes de Dynkin }}}}}{\underline{\centerline{\textit{\small{
Matrices de Cartan et diagrammes de Dynkin}}}}}

Maintenant que la  notion  de racines  simples  est  mise en place,
on peut donner la classification des alg\`ebres de Lie.
Premi\`erement, nous   allons  parler  d'une  une matrice qui donne
un renseignement sur la longueur des racines et l'angle entre elles.
Par la suite, nous pr\'esentons  un diagramme permettant de
retrouver les m\^emes donn\'es.

\section{Matrice de Cartan}
\subsection{D\'efinition}
Soit $g$ une alg\`ebre de Lie de dimensions finie.  On appelle la matrice de Cartan  associ\'ee
\`a  $\Delta $ relativement \`{a} $\pi,  $ la matrice $K$ de coefficients
\begin{eqnarray*}
K_{ij}=<\alpha _{i},\alpha _{j}>=2\frac{\alpha _{i}.\alpha
_{j}}{\alpha _{j}.\alpha _{j}}
 \end{eqnarray*} o\`u  les
$\alpha_i$ sont des racines simples.\\
{\bf Remarques:}
\begin{itemize}
\item $K$ une matrice carr\'{e}e
$\ell \times \ell$,  o\'u  $\ell$  est le rang de $g$.
\item Il y a seulement les possibilit\'es suivantes
\begin{eqnarray*}
K_{ii}&=&2,\quad \forall i, \; i=1,\ldots,\ell \\
K_{ij}&=&0,-1,-2,-3, \quad i \neq j.
\end{eqnarray*}
\end{itemize}
\subsection{Exemples}
Nous donnons quelque exemples. \\
{\bf Exemple 1:}. Pour l'alg\`ebre $A_1$, la matrice de Cartan   se
r\'eduit \`a
\begin{eqnarray*}
K=2. \end{eqnarray*} {\bf Exemple 2:}.  Pr\'esentons  les alg\`ebres
ayant le rang  2.  En effet, nous avons  quatres possibilit\'es \\1.
Alg\`ebre de Lie $A_1\oplus A_1$.   Sa matrice de Cartan est comme
suit
\begin{eqnarray*}
 K=\left(
\begin{array}{cc}
2 & 0 \\
0 & 2%
\end{array}%
\right).
\end{eqnarray*}
2. La matrice de Catan associ\'ee a $A_2$ est donn\'ee par
\begin{eqnarray*}
 K=\left(
\begin{array}{cc}
2 & -1\\
-1 & 2%
\end{array}%
\right)
\end{eqnarray*}
3. La matrice de Cartan associ\'ee  \`a $B_2$ est comme suit
\begin{eqnarray*}
K=\left(
\begin{array}{cc}
2 & -2 \\
-1 & 2%
\end{array}%
\right).
\end{eqnarray*}
4. On d\'efinit l'alg\`ebre de Lie $G_2$ avec  la matrice de Cartan
suivante:
\begin{eqnarray*}
K=\left(
\begin{array}{cc}
2 & -1 \\
-3 & 2%
\end{array}%
\right).
\end{eqnarray*}
\subsection{Propri\'{e}t\'{e}s de  la matrice  de Cartan}
Soit $K$  une matrice de Cartan
\begin{center}
$K=\left(
\begin{array}{ccc}
\left\langle \alpha _{1},\alpha _{1}\right\rangle & \left\langle
\alpha
_{1},\alpha _{2}\right\rangle & \ldots \\
\left\langle \alpha _{2,}\alpha _{1}\right\rangle & \left\langle
\alpha _{2},\alpha _{2}\right\rangle& \ldots\\
\ldots& \ldots& \ldots
\end{array}%
\right). $
\end{center}
Nous avons les propri\'{e}t\'{e}s suivants: \\$\bullet $ Les
entr\`{e}s $K_{ij}\in Z $.\\
 $\bullet $ $K$ d\'epend de l'ordre des
racines $\alpha _{i}$ de $\pi $.\\ $\bullet $ $K$ ne d\'epend pas de
la base $\pi $.\\ $\bullet$  La matrice de Cartan est non
d\'eg\'en\'er\'ee si  $det(A) \neq 0$.
\subsubsection{Proposition:}
Soient $\left( \Delta ,E\right) $ et $\left( \Delta ^{\prime
},E^{\prime }\right) $ deux syst\`{e}mes des racines \ ayant pour
base $\pi =\left\{ \alpha _{i}\right\} $ et $\pi =\left\{ \beta
_{i}\right\} $ respectivement. Si la  matrice de Cartan  $K$
coincide avec   $K'$
\begin{eqnarray*} \left\langle \alpha _{i},\alpha
_{j}\right\rangle =\left\langle \beta _{i},\beta _{j}\right\rangle
\end{eqnarray*}
alors\\
 $\bullet $ La bijection $\Phi :\alpha _{i}\to \beta
_{i}$ se prolonge de fa\c{c}on unique en isomorphisme $\Phi: E\to
E^{\prime }$
\begin{eqnarray*} \Phi \left( \Delta \right) =\Delta
^{\prime }
\end{eqnarray*}
et
\begin{eqnarray*} \left\langle \Phi \left( \alpha \right) ,\Phi
\left( \beta \right) \right\rangle =\left\langle \alpha ,\beta
\right\rangle,\;\;\;\;\; \forall \alpha ,\beta \in \Delta.
\end{eqnarray*}
$\bullet $ La matrice de Cartan $K$ d\'etermine le syst\`{e}me des racines de fa%
\c{c}on unique \`{a} un isomorphisme pr\'{e}s.
\section{Graphes de Coxter et diagrammes \ de Dynkin}

\section{Graphes de Coxter}
Le but de   diagramme de  Coxter  est de r\'ecapituler les
informations de la matrice de Cartan en un diagramme.
 On appelle
graphe de Coxter de $\Delta $ relativement \`{a} $\pi $ le graphe
dont les vertex sont les \'el\'ements de $\pi $  et les vertex
$\alpha _{i}$ et $\alpha _{j}$ $\left( i\neq j\right) $ sont joints
par $\left\langle \alpha _{i},\alpha _{j}\right\rangle \left\langle
\alpha _{j},\alpha _{i}\right\rangle $ lignes o\`u $ 0\leq
\left\langle \alpha _{i},\alpha _{j}\right\rangle \left\langle
\alpha _{j},\alpha _{i}\right\rangle \leq 4$.  Les veleurs possibles
de  $ \left\langle \alpha _{i},\alpha _{j}\right\rangle \left\langle
\alpha _{j},\alpha _{i}\right\rangle$ sont $\left\{
0,1,2,3,4\right\} $. \\ Ce praphe est  repr\'esent\'e  comme suit
\begin{itemize}
\item  Les racines $\alpha_i$ sont repr\'esent\'ees par des vertex.
\item L'angle entre les racines est repr\'esent\'e par le nombre de lignes, reliant ces deux racines.
\end{itemize}
{\bf Remarque:} Si $\alpha _{i}$ et $\alpha _{j}$ sont deux racines
simples tel que $\ \ \ $
\begin{center}
$\left\vert \alpha _{i}\right\vert =\left\vert \alpha
_{j}\right\vert $
\end{center}
alors $K$ est sym\'{e}trique
\begin{center}
$K_{ij}=K_{ji}$.
\end{center}
Cette condition  permet de classifier  les alg\`{e}bres de Lie
simplement lac\'{e}es, $ADE$. La condition  $K_{ij}\neq K_{ij}$
d\'etermine  les alg\`{e}bre de Lie non simplement lac\'{e}es $%
C_{n},B_{n},F_{4},G_{2}$ o\`u $\left\vert \alpha _{i}\right\vert
\neq \left\vert \alpha _{j}\right\vert $.
\subsection{Diagrammes de Dynkin}
En r\'ealit\'{e} le graphe de Coxter ne suffit pas \`{a}
d\'{e}terminer la matrice de Cartan, il ne fournit que les angles
entre les racines simples, sans indiquer laquelle est la plus
longue. Il faut remplacer les diagramme de Coxter par des diagrames
ori\'{e}nt\'{e}s.
\subsubsection*{D\'{e}finition:}
C'est un graphe de Coxter  (ori\'{e}nt\'{e}) qui dans le cas ou 2 vertex $i$ et $j$ ont une fl\`{e}%
che partant de la racine longue vers la racine courte $\left\vert
\alpha _{i}\right\vert \geq \left\Vert \alpha _{j}\right\Vert .$

 Nous pr\'esentons des exemples:\\
 {\bf Exemple 1:  Le diagramme de Dynkin de $A_1$}.\\   Rappelons que  le rang de $A_1$ est  1, on n'a qu'une
racine simple $\alpha$. Son digramme  est comme suit
\begin{eqnarray*}
\mbox{
         \begin{picture}(20,30)(70,0)
        \unitlength=2cm
        \thicklines
    \put(0,0.2){\circle{.2}}
     \put(-1.2,.15){$A_{1}:$}
  \end{picture}
} \label{ordA1}
\end{eqnarray*}
{\bf Exemple 2:  Le diagramme de Dynkin de $A_1\oplus A_1$}.\\
C'est un  diagrammes de Dynkin \`a deux  vertex s\'epar\'es:
\begin{eqnarray*}
\mbox{
         \begin{picture}(20,30)(70,0)
        \unitlength=2cm
        \thicklines
     \put(2.5,0.2){\circle{.2}}
     \put(3.2,0.2){\circle{.2}}
     \put(-1.2,.15){$A_1\oplus A_1$:}
  \end{picture}
} \label{ordA2}
\end{eqnarray*}
{\bf Exemple 3:  Le diagramme de Dynkin de $A_2$}.\\
C'est un  diagramme de Dynkin \`a deux  vertex connect\'es par une seule ligne:
\begin{eqnarray*}
\mbox{
         \begin{picture}(20,30)(70,0)
        \unitlength=2cm
        \thicklines
     \put(2.5,0.2){\circle{.2}}
     \put(2.6,0.2){\line(1,0){.5}}
     \put(3.2,0.2){\circle{.2}}
     \put(-1.2,.15){$A_{2}:$}
  \end{picture}
} \label{ordA2}
\end{eqnarray*}
{\bf Exemple 4:  Le diagramme de Dynkin de $D_4$}.\\
C'est un  diagrammes de Dynkin qui contient  un  vertex trivalent:
\begin{eqnarray*}
 \mbox{
         \begin{picture}(20,100)(90,-50)
        \unitlength=2cm
        \thicklines
      \put(-1.2,-.05){$D_4:$}
\put(2.5,0){\circle{.2}}
     \put(2.6,0){\line(1,0){.5}}
     \put(3.2,0){\circle{.2}}
     \put(3.2,.1){\line(1,1){.35}}
     \put(3.64,.5){\circle{.2}}
     \put(3.2,-.1){\line(1,-1){.35}}
     \put(3.64,-.5){\circle{.2}}
  \end{picture}
} \label{ordDk}
\end{eqnarray*}
Les diagrammes de Dynkin permettent de classifier     les alg\`ebres de Lie.  Rappelons la classification $ADE$:
\begin{enumerate}
\item Alg\`ebres de Lie $A_n$:  isomorphe \`a   $sl(n + 1,C)$,  ($n\geq 1$)
\item  Alg\`ebres de Lie $B_n$: isomorphe  \`a  $so(2n + 1,C)$, ($n\geq 2$)
\item Alg\`ebres de Lie $D_n$:  isomorphe \`a  $so(2n)$,  ($n\geq 4$)
\item Alg\`ebres
exceptionnelles: $E_6$, $E_7$, $E_8$, $F_4$ et $G_2$.
\end{enumerate}
Les diagrammes de Dynkin de $ADE$   sont class\'es comme suit
\begin{eqnarray*}
 \mbox{
         \begin{picture}(20,30)(70,0)
        \unitlength=2cm
        \thicklines
    \put(0,0.2){\circle{.2}}
     \put(.1,0.2){\line(1,0){.5}}
     \put(.7,0.2){\circle{.2}}
     \put(.8,0.2){\line(1,0){.5}}
     \put(1.4,0.2){\circle{.2}}
     \put(1.6,0.2){$.\ .\ .\ .\ .\ .$}
     \put(2.5,0.2){\circle{.2}}
     \put(2.6,0.2){\line(1,0){.5}}
     \put(3.2,0.2){\circle{.2}}
     \put(-1.2,.15){$A_{n}:$}
  \end{picture}
} \label{ordAk}
\end{eqnarray*}
\begin{eqnarray*}
 \mbox{
         \begin{picture}(20,100)(90,-50)
        \unitlength=2cm
        \thicklines
      \put(-1.2,-.05){$D_n:$}
    \put(0,0){\circle{.2}}
     \put(.1,0){\line(1,0){.5}}
     \put(.7,0){\circle{.2}}
     \put(.8,0){\line(1,0){.5}}
     \put(1.4,0){\circle{.2}}
     \put(1.6,0){$.\ .\ .\ .\ .\ .$}
     \put(2.5,0){\circle{.2}}
     \put(2.6,0){\line(1,0){.5}}
     \put(3.2,0){\circle{.2}}
     \put(3.2,.1){\line(1,1){.35}}
     \put(3.64,.5){\circle{.2}}
     \put(3.2,-.1){\line(1,-1){.35}}
     \put(3.64,-.5){\circle{.2}}
  \end{picture}
} \label{ordDk}
\end{eqnarray*}
\begin{eqnarray*}
\mbox{
         \begin{picture}(20,350)(50,-20)
        \unitlength=2cm
        \thicklines
    \put(0,4){\circle{.2}}
     \put(.1,4){\line(1,0){.5}}
     \put(.7,4){\circle{.2}}
     \put(.8,4){\line(1,0){.5}}
     \put(1.4,4){\circle{.2}}
    \put(1.5,4){\line(1,0){.5}}
    \put(2.1,4){\circle{.2}}
    \put(2.2,4){\line(1,0){.5}}
    \put(2.8,4){\circle{.2}}
   \put(1.4,4.1){\line(0,1){.5}}
   \put(1.4,4.7){\circle{.2}}
   \put(-.7,2){\circle{.2}}
   \put(-.6,2){\line(1,0){.5}}
    \put(0,2){\circle{.2}}
     \put(.1,2){\line(1,0){.5}}
     \put(.7,2){\circle{.2}}
     \put(.8,2){\line(1,0){.5}}
     \put(1.4,2){\circle{.2}}
    \put(1.5,2){\line(1,0){.5}}
    \put(2.1,2){\circle{.2}}
    \put(2.2,2){\line(1,0){.5}}
    \put(2.8,2){\circle{.2}}
   \put(1.4,2.1){\line(0,1){.5}}
   \put(1.4,2.7){\circle{.2}}
  \put(-1.4,0){\circle{.2}}
  \put(-1.3,0){\line(1,0){.5}}
  \put(-.7,0){\circle{.2}}
   \put(-.6,0){\line(1,0){.5}}
    \put(0,0){\circle{.2}}
     \put(.1,0){\line(1,0){.5}}
     \put(.7,0){\circle{.2}}
     \put(.8,0){\line(1,0){.5}}
     \put(1.4,0){\circle{.2}}
    \put(1.5,0){\line(1,0){.5}}
    \put(2.1,0){\circle{.2}}
    \put(2.2,0){\line(1,0){.5}}
    \put(2.8,0){\circle{.2}}
   \put(1.4,.1){\line(0,1){.5}}
   \put(1.4,.7){\circle{.2}}
   \put(-2.3,.3){${E}_8:$}
   \put(-2.3,2.3){${E}_7:$}
   \put(-2.3,4.7){${E}_6:$}
  \end{picture}}
\label{affE}
\end{eqnarray*}

\chapter{   G\'en\'eralit\'es sur  la th\'eorie des groupes}
\pagestyle{myheadings}
\markboth{\underline{\centerline{\textit{\small{  Generalites sur
les groupes}}}}}{\underline{\centerline{\textit{\small{ Generalites
sur les groupes}}}}} Dans ce chapitre, nous allons donner des
notions sur des la th\'eeorie des  groupes. A traver des exemples,
nous allons traiter plus particuli\`erement  les trois cat\'egories:
\begin{enumerate}
\item  Groupes finis
\item Groupes discrets infinis
\item Groupes continus.
\end{enumerate}
\section{Structure du groupe}
\textbf{D\'{e}finition:} Un groupe $G$ est un ensemble
d'\'{e}l\'{e}ments   muni d'une loi de   composition telle que:
\\$\bullet $ La loi de composition est interne
\begin{eqnarray*}
g_{1}.g_{2}\in G,\qquad \forall g_{1},g_{2}\in G
\end{eqnarray*}
 $\bullet $ Il existe un
\'{e}l\'{e}ment neutre
\begin{eqnarray*}
e.g=g.e=g, \qquad \forall  g\in G
\end{eqnarray*} $\bullet $ Chaque \'{e}l\'{e}ment poss\`{e}de un
inverse tel que
\begin{eqnarray*}
g.g^{-1}=g^{-1}.g=e
\end{eqnarray*}
$\bullet $ La loi de composition est associative
\begin{eqnarray*}
g_{1}.(g_{2}.g_{3})=(g_{1}.g_{2}).g_{3}.
\end{eqnarray*}
{\bf Remarques:}
\begin{enumerate}
\item
  La loi  de composition et les
\'{e}l\'{e}ments d\'{e}ffinissent un groupe.
\item Ces deux
informations peuvent \^{e}tre repr\'{e}sent\'{e}s dans une table de
multiplication
\begin{table}[th]
\begin{center}
\begin{tabular}{|c|c|c|c|c|c|c|c|c|c|c|}\hline
& $e$ & $g_{1}$ & $g_{2}$ & $g_{3}$ & ............ \\\hline $e$ &
$e$ & $g_{1}$ & $g_{2}$ & $g_{3}$ & ............ \\ \hline $g_{1}$ &
$g_{1}$ & $g_{1}.g_{1}$ & $g_{1}g_{2}$ & $g_{1}g_{3}$ & ............ \\
\hline $g_{2}$
& $g_{2}$ & $g_{2}.g_{1}$ & $g_{2}.g_{2}$ & $g_{2}.g_{3}$ & ............ \\
\hline $g_{3}$ & $g_{3}$ & $g_{3}.g_{1}$ & $g_{3}.g_{2}$ &
$g_{3}.g_{3}$ &
............ \\ \hline... & ... & ... & ... & ... & ............\\
\hline
\end{tabular}%
\caption{La table de multiplication} \label{t:one}
\end{center}
\end{table}
\item  La table de multiplication donne just des informations
sur la structure abstraite du groupe.\item  Un groupe peut
\^{e}tre ab\'{e}lien ou non ab\'{e}lien (commutatif, non
commutatif).
\end{enumerate}
\textbf{Proposition:} \\Un groupe $G$ est dit
ab\'{e}lien si et  seulement si
\begin{eqnarray*}
g_{1}.g_{2}=g_{2}.g_{1}, \qquad \forall   g_{1},g_{2}\in G.
 \end{eqnarray*} Dans ce cas, la table de
multiplication est sym\'{e}trique.\\
En g\'en\'erale,  les  groupes se divisent en  trois cat\'{e}gories des groupes:
\begin{enumerate}
\item
\textbf{Groupes finis}\footnote{Nombre finid'\'{e}l\'{e}ments}.\item
\textbf{Groupes discret infinis}\footnote{Nombre infini d'\'{e}l\'{e}%
ments}.\item \textbf{Groupes continus.}\footnote{(nombre infinis
d'\'{e}l\'{e}ments) et chaque \'el\'ement est defini \`a travers des
param\`{e}tres $ g=f(\phi _{1}.......\phi _{n})$,  o\`u  $n$ le
nombre des param\`{e}tres continus: la dimension du groupe.}
\end{enumerate}

\section{Exemples}
Dans cette section, nous donnons quelques exemples.
\subsection{Groupes finis}
Les groupes finis sont les plus simples.  Le nombre
d'\'{e}l\'{e}ment d'un groupe fini s'appelle l'ordre du groupe.  On
la note $\left\vert G\right\vert $. \par  $\bullet $ Par exemple,
l'ensemble qui contient les  deux nombres $+1$ et $-1$, muni de la
multiplication habituelle, forme un groupe ab\'{e}lien.
Sa table de multiplication est donn\'{e}e par
\begin{table}[th]
\begin{center}
\begin{tabular}{|c|c|c|}\hline
& 1 & -1 \\ \hline 1 & 1 & -1 \\ \hline -1 & -1 & 1 \\ \hline
\end{tabular}
\caption{La table de multiplication de (1,-1)} \label{t:two}
\end{center}
\end{table}

$\bullet $ Groupes cycliques  $Z_{n}$\\
Le groupe $Z_{n}$  est d\'{e}fini par $Z _{n}=\left\{ g^{k},1\leq
k\leq n,g^{n}=e\right\} $. L'ordre de $Z_{n}$ est $n$. Tous les
\'el\'ements s'\'ecrivent comme puissances de l'un d'entre eux.   Ce
groupe est  ab\'{e}lien.\newline Dans le cas $n=2$, le groupe
devient $Z_{2}$ et sa table de multiplication est donn\'ee in
(\ref{t:three}).
\begin{table}[th]
\begin{center}
\begin{tabular}{|c|c|c|}\hline
& e & g \\ \hline e & e & g \\ \hline
g & g & e%
\\ \hline
\end{tabular}
\caption{La table de multiplication de $Z_{2}$} \label{t:three}
\end{center}
\end{table}

$\bullet $ {Groupes sym\'{e}triques $S_{n}$}\\  Rappelons que  les
groupes sym\'{e}triques ont une importance non n\'egligeable dans
certains domaines de la physique quantique. Dans   un ensemble de
$n$ objets,  il y a $n!$ permutations  possibles  qui   d\'eplacent les
$n$ objets d'une mani\`ere  arbitraire.   Ce groupe de permutations
est not\'e par  $S_{n}$ et son ordre    est
\begin{eqnarray*}
\left\vert S_{n}\right\vert =n! \end{eqnarray*}
\textbf{Remarques:}\\
- Pour $n=2$,
$S_{2}$ devient $Z_2$ et  il est ab\'elien. \\
- Pour $n\geq 3$,    $S_{n}$ est non ab\'elien.

\subsection{Groupes discrets infinis}
Les exemples de groupes discrets infinis sont nombreux. Nous citons
par exemple:\\
$\bullet $ L'exemple le plus simple  est $(Z,+)$.\\
$\bullet $ Le groupe $SL(2,Z)$: le groupe des matrices $2\times 2$
\`{a} co\'{e}fficients dans $Z$ et le d\'{e}terminant
 $1$.  Rappelons que les \'el\'ements de $SL(2,Z)$ sont des matrices
\begin{eqnarray*}
M= \left(
\begin{array}{cc}
a & b\\
c & d%
\end{array}%
\right)
\end{eqnarray*}
avec la condition $ad-bc=1$, o\`u    $ a,b,c,d \in  Z$.

\subsection{Groupes continus}
En physique, les groupes continus sont appel\'{e}s groupe de Lie.
Dans ce
cas,  les \'{e}l\'{e}ments du groupe d\'{e}pendent  des param\`{e}tres r\'{e}%
els continus ind\'ependants. Le nombre de ces param\`{e}tres est
appel\'{e} la dimension du groupe. En effet, on a
\begin{eqnarray*}
g=f(\theta _{1}........\theta _{n}),
\end{eqnarray*}
o\`u $n$ est la dimension dy groupe.\\
\textbf{Exemples:} \\
- Le groupe de rotations de l'espace \`{a} $2$ dimension $SO(2)$. Sa
dimension est 1 et chaque rotation est donn\'e par une matrice de la
forme suivante
\begin{eqnarray*}
R= \left(
\begin{array}{cc}
cos\;\theta & -sin\;\theta \\
sin\;\theta &  cos\;\theta
\end{array}%
\right). \end{eqnarray*} - Le groupe de rotation $SO(n)$ de l'espace
euclidien \`a  $n$ dimensions g\'en\'eralisant  les rotations de
\`a deux dimensions. Sa dimension est $ \frac{n(n-1)}{2}$.
\section{Classe d'un groupe }
 Soit $G$ un groupe. On d\'{e}finit sur  $G$ la relation d'\'{e}quivalence
suivante:
\begin{eqnarray*}
a\sim b,\qquad  si \;\;\;\exists  g\in G,\;\;\;a=g.b.g^{-1}
\end{eqnarray*} \textbf{Remarques:}\\
- On dit aussi les \'{e}l\'{e}ments $a$ et $b$ sont conjugu\'{e}s.\\
- On  remarque  que $a$ est conjugu\'{e} de lui m\^{e}me
$a=e.a.e^{-1}$.\\- Si $a\sim b$ et $b\sim c$,  alors $a\sim c$. En
effet, on a
\begin{eqnarray*}
a\sim b\Longrightarrow  tel que  a=g.b.g^{-1}=g.g^{\prime
}.c.g^{^{\prime }-1}g^{-1}=(g.g^{^{\prime }}).c.(gg^{^{\prime
}})^{-1}. \end{eqnarray*}
- Les classes d'\'{e}quivalence r\'{e}alisent une partition de $G$%
\begin{eqnarray*}
G=\left\{ \left\{ e\right\} ,\left\{ ..\right\},\ldots\right\}.
\end{eqnarray*}
\section{Sous groupes}
\subsection{D\'efinitions}
\textbf{D\'{e}finition 1:} Un sous ensemble non vide $H$ d'un groupe
$G$ est dit un sous groupe si la restriction de la loi du $G$ \`{a}
$H$ d\'{e}termine sur $H$ une loi du  groupe.\\
\textbf{D\'{e}finition 2:}\\
Un sous ensemble $H$ d'un groupe $G$ est dit un sous groupe de $G$
si et seulement si:
\begin{eqnarray*}
H &\neq &\phi  \\
\forall x,y &\in &H, \quad x.y^{-1}\in H.
\end{eqnarray*}
\textbf{D\'{e}monstration:}  Puisque $H$ est un sous groupe, $H$ est
non vide  $(e\in H).$\\
Soir $h$ un \'el\'ement de $H$, alors nous avons $
y^{-1} \in  H, \; \forall  y\in H$.\\
Pour tout \'el\'ement $x$  de  $H$, on a aussi $x.y^{-1} \in  H$
donc $H$ est stable par l'inverse et la loi du groupe.\\
Soit $H$ un sous ensemble tel que  $ \forall  x,y\in H, \;\;
x.y^{-1}\in H$, on a
\begin{eqnarray*}
y &=&x\Longrightarrow x.x^{-1}=e\in H \\
x &=&e,y=x\Longrightarrow e.x^{-1}\in H\Longrightarrow x^{-1}\in H
\end{eqnarray*}%
 donc
 \begin{eqnarray*}
x.(y^{-1})^{-1}\in H\Longrightarrow x.y\in H.
\end{eqnarray*}\\
\textbf{Proposition}:  Si $H$ est un sous groupe de $G$ fini, alors
on a
\begin{eqnarray*}
\left\vert H\right\vert \leq \left\vert G\right\vert.
\end{eqnarray*}
\subsection{Exemples}
Nous donnons quelquels exemples.\newline \textbf{Exemple 1:}
L'\'{e}l\'{e}ment neutre $\left\{ e\right\} $ est un sous groupe de
$G$. Puisque nous avons $ \left\{ e\right\} \neq \phi$ et
$e.e^{-1}=e.e=e\in \left\{ e\right\}$. Le sous groupe  $\left\{
e\right\} $ est un sous groupe trivial.\\
\\
\textbf{Remarque:}\ Tout sous groupe diff\'erent de $\left\{
e\right\} $ et $G$
est dit propre.\\ \textbf{Exemple 2:} Le centre du groupe\\
On d\'efinit le centre de $G$ comme suit
\begin{eqnarray*}
Z(G)=\left\{ h\in G| \;\; hg=gh, \forall  g\in G\right\}.
\end{eqnarray*}
$Z(G)$ est un sous groupe de $G$ ab\'{e}lien.\\
{\bf D\'emonstration}\\
Soient $h_1$ et $h_2$ deux \'el\'ements de $Z(G)$, alors nous avons
\begin{eqnarray*}
e\in Z(G)\Longrightarrow &&e.g=g.e\\
h_{1},h_{2} &\in &Z(G)\Longrightarrow h_{1}.h_{2}\in Z(G) \\
h_{1}h_{2}.g &=&g.h_{1}h_{2} \\
(gh)^{-1} &=&g^{-1}.h^{-1}=h^{-1}.g^{-1} \\
g &\in &G, g^{-1}\in G\Longrightarrow h^{-1}\in Z(G).
\end{eqnarray*}
\textbf{Exemple 3 :} Le groupe de sym\'{e}trie d'un triangle
\'{e}quilat\'{e}ral. \\ Ce groupe not\'e  $D_{3}$ poss\`{e}de six
\'{e}l\'{e}ments. Ils sont class\'{e}s comme suit:\\
$\bullet $ Trois r\'{e}flexions $\sigma _{1}$, $\sigma_{2}$,
$\sigma_{3}$.\\
$\bullet $ Trois rotations autour de  l'axe $z$.  Elles sont
determin\'{e}es par les rotations d'angles: $ 0, \frac{2\pi
}{3},\frac{4\pi }{3}=-\frac{2\pi }{3}$, (avec 0  est l'\'el\'ement
neutre).\\
On note ces  \'{e}l\'{e}ments par  $\left\{ e,C_{3},C_{-3} \sigma
_{1,}\sigma _{2},\sigma _{3}\right\} $. Son ordre est donc
$\left\vert D_{3}\right\vert =6$ $\langle $ $\infty $ \ ( donc il
est fini).  Ce groupe  n'est pas ab\'{e}lien puisque $\sigma
_{2}C_{3}=\sigma _{3}\neq C_{3}\sigma _{2}=\sigma _{1}$.
\begin{table}[th]
\begin{center}
\begin{tabular}{|c|c|c|c|c|c|c|c|c|c|c|}\hline
& $e$ & $\sigma _{1}$ & $\sigma _{2}$ & $\sigma _{3}$ & $C_{3}$ & $C_{-3}$ \\
\hline $e$ & $e$ & $\sigma _{1}$ & $\sigma _{2}$ & $\sigma _{3}$ &
$C_{3}$ & $C_{-3}$ \\ \hline $\sigma _{1}$ & $\sigma _{1}$ & $e$ &
$C_{3}$ & $C_{-3}$ &$\sigma _{2}$ &$\sigma _{3}$
\\ \hline
$\sigma _{2}$ &  $\sigma_2$& $C_3$ & $e$ & $C_{3}$ & $\sigma_3$ & $\sigma_1$ \\
\hline $\sigma _{3}$ & $\sigma _{3}$&$C_{3}$ &$C_{-3}$ & $e$ &
$\sigma_1$  & $\sigma_2$
\\ \hline
$C_{3}$ & $C_{3}$ & $\sigma_3$ & $\sigma_1$ & $\sigma_2$ & $C_{-3}$  & $e$ \\
\hline $C_{-3}$ & $C_{-3}$  & $\sigma_2$& $\sigma_3$ &  $\sigma_1$&
$e$ &$C_{3}$
\\
\hline
\end{tabular}%
\caption{La table de multiplication de $D_{3}$ } \label{12}
\end{center}
\end{table}

 Les sous
groupes de $D_{3}$ sont class\'{e}s comme suit:
\begin{enumerate}
\item L'\'{e}lements neutre $\left\{ e\right\} $ : un sous groupe trivial d'ordre 1.
\item $\left\{ e,\sigma _{1}\right\} $:  un sous groupe propre d'ordre 2
\item $\left\{ e,\sigma _{2}\right\} $:  un sous groupe propre d'ordre 2
\item $\left\{ e,\sigma _{3}\right\} $:  un sous groupe propre d'ordre 2
\item $\left\{ e,\frac{2\Pi }{3},\frac{-2\Pi }{3}\right\} $:  un sous groupe
propre d'ordre 3.  (avec $C_{3}\equiv\frac{2\Pi }{3},
C_{-3}\equiv\frac{-2\Pi }{3}$).
\end{enumerate}

\textbf{D\'{e}finition:}

Un sous groupe $H$ est dit invariant si et seulement si:
\begin{eqnarray*}
g.h.g^{-1}\in H, \forall  h\in H et g\in G.
\end{eqnarray*}

\textbf{Exemple:} Le centre de $G$ est un sous groupe invariant
\begin{eqnarray*}
\forall h\in Z(G), \quad g.h^{-1}.g^{-1}=g.g^{-1}.h^{-1}=h^{-1}\in
H.
\end{eqnarray*}

\section{Homomorphismes }

On consid\`{e}re deux groupes $G$ et $G^{\prime }$ et une application $%
\varphi $ de $G$ dans $G^{\prime }$ . On dit que $\varphi $ est un
homomorphisme si nous avons
\begin{eqnarray*}
\forall (g_{1},g_{2})\in G, \;\;\varphi (g_{1}.g_{2})=\varphi
(g_{1}).\varphi (g_{2}).
\end{eqnarray*}
Si $\varphi $ un homomorphismes, alors \\
- $\varphi (e)$  est l'\'{e}l\'{e}ment neutre de $G^{\prime
}$.\\
- $\varphi (g^{-1})$  est l'inverse de $\varphi (g)$.\\
\textbf{D\'{e}finition:}  Le noyau d'un homomorphisme est d\'{e}fini
par
\begin{eqnarray*}
Ker \varphi =\left\{ g\in G|\;\; \varphi (g)=e\right\}
\end{eqnarray*}
 Pour terminer ce chapitre, nous presentons quelques  propositions:
\begin{enumerate}
\item $Ker$ $\varphi $ est un sous groupe invariant de $G$. Il faut
montrer  si $h\in Ker$ $\varphi $, alors $ g.h.g^{-1}\in Ker\varphi,
\forall g\in G $. En effet,
\begin{eqnarray*}
\varphi (g.h.g^{-1}) &=&\varphi (g)\varphi (h)\varphi (g^{-1}) \\
&=&\varphi (g)\varphi (g^{-1}) \\
&=&\varphi (gg^{-1}) \\
&=&\varphi (e) \\
&=&e
\end{eqnarray*}
et par cons\'equent  $g.h.g^{-1}\in Ker$ $\varphi $.\\
 \item Si $\varphi $ est bijectif, $\mathbf{\varphi }$
est un homomorphisme.
\end{enumerate}

\chapter{   Repr\'esentations des Groupes}
\pagestyle{myheadings}
\markboth{\underline{\centerline{\textit{\small{ Repr\'esentations
des Groupes:
G\'een\'eralit\'es}}}}}{\underline{\centerline{\textit{\small{
Repr\'esentations des Groupes: G\'en\'eeralit\'es}}}}}

\section{ D\'{e}finitions et exemples}
En physique, si  une th\'eorie  poss\`ede une invariance sous un
groupe,  il faut   savoir comment le groupe agit sur les variables
et les  grandeurs physiques en utilisant des  repr\'esentations.
\subsection{Repr\'esentations}
 Consid\'{e}rons un groupe  d'\'{e}l\'{e}ments ($G=\left\{
e,g,...\right\} $).  Une repr\'{e}sentation de $G$ est un ensemble
des matrices $D(g)$ (une matrice pour chaque \'{e}l\'{e}ment), qui
se composent exactement comme les \'{e}l\'{e}ments du groupe. Ces
matrices satisfont les relations suivantes
\begin{eqnarray*}
D\left( g_{1}g_{2}\right) &=&D\left( g_{1}\right) D\left(
g_{2}\right)\\
D\left( e\right) &=&I_{n\times n}\\
D\left( g^{-1}\right) &=&\left( D\left( g\right) \right) ^{-1}.
\end{eqnarray*}
{\bf Remarques}:
\begin{enumerate}
\item  Une repr\'{e}sentation est un homomorphisme.
\item  Pour chaque $g$, $D(g)$   est une matrice $n\times n$, o\`u $n$ est la dimension de la
repr\'{e}sentation:
\begin{eqnarray*}
D\left( g\right) =\left(
\begin{array}{ccc}
&  \\ matrice & \\
&
\end{array}%
\right) _{n\times n}.
\end{eqnarray*}
\item  Les matrices $D(g)$ sont suppos\'{e}es  inversibles.
\item  Si $G$ est d\'{e}fini comme un groupe des matrices, ($SO(n)$ par
exemple), sa d\'efinition est donn\'ee en terms  d'une
repr\'esentation. \end{enumerate}
 {\bf Remarque: R\'epresentation triviale}\\ On peut associer
\`a tout \'{e}l\'{e}ment $g$  la matrice  identit\'e $I_{n\times n}$
( la  matrice carr\'ee avec des 1 sur la diagonale et des 0 partout
ailleurs).  Cette repr\'{e}sentation est dite \emph{\ trivale.}

\subsection{Repr\'{e}sentation fid\'{e}le}
Lorsque $g_{1}\neq g_{2}$ et $D\left( g_{1}\right) \neq D\left(
g_{2}\right) $,  la repr\'{e}sentation est dite \emph{fid\'{e}le}.
En effet, on a
\begin{eqnarray*} g_{1}\neq g_{2}\Longleftrightarrow
D\left( g_{1}\right) \neq D\left( g_{2}\right).
\end{eqnarray*}
Dans cette repr\'esentation, la  correspondance  ($ G \to
matrices$)
\begin{eqnarray*}
\left(
\begin{array}{c}
e \\
g_{1} \\
. \\
. \\
. \\
g_{n}%
\end{array}%
\right) \longmapsto \left(
\begin{array}{c}
I_{n\times n} \\
D\left( g_{1}\right)  \\
. \\
. \\
. \\
D\left( g_{n}\right)
\end{array}%
\right)
\end{eqnarray*}
est un isomorphisme.\\
\\ {\bf  Exemples}:
\begin{enumerate}
\item  Notre premier exemple   simple   est $G=\{e\}$. La structure du groupe est donne\'{e} par  $e\cdot
e=e$. Alors on peut  repr\'esenter ce groupe par une
repr\'esentation de dimension $1$ comme suit $ e =1$. Une
repr\'{e}sentation de dimention 2  est donn\'ee par
\begin{eqnarray*}
e=\left(
\begin{array}{cc}
1 & 0 \\
0 & 1%
\end{array}%
\right).
\end{eqnarray*}
\item Le deuxi\`eme  exemple est  le   groupe cyclique not\'e $Z_{2}=\left\{e,\;g\right\}$ tel que
$g\cdot g=g^{2}=e$.  Sa structure est comme suit
\begin{eqnarray*}
\ e\cdot e&=&e\\
 e\cdot g&=& g\cdot e=g.\\
g\cdot g&=&e
\end{eqnarray*}
Parmis les repr\'esentations de $Z_{2}$, on cite:
\begin{itemize}
\item Repr\'{e}sentation triviale de dimension 1%
\begin{eqnarray*}
e=1, \qquad  g=1
\end{eqnarray*}
 \item Repr\'{e}sentation triviale de dimension 2
\begin{eqnarray*}
e=\left(
\begin{array}{cc}
1 & 0 \\
0 & 1%
\end{array}%
\right),\qquad g=\left(
\begin{array}{cc}
1 & 0 \\
0 & 1%
\end{array}%
\right)
\end{eqnarray*}
\item Repr\'{e}sentation  non triviale et fid\'{e}le  de dimension  1
\begin{eqnarray*}
e=1,\qquad  g=-1
\end{eqnarray*}
\item  Repr\'{e}sentation non triviale et  fid\'{e}le  de dimension
2
\begin{eqnarray*}
 e=\left(
\begin{array}{cc}
1 & 0 \\
0 & 1%
\end{array}%
\right),\qquad
  g=\left(
\begin{array}{cc}
1 & 0 \\
0 & -1%
\end{array}%
\right)
\end{eqnarray*}
\end{itemize}
\item Le  groupe de Lie $SO(2)$ est donn\'e par
\begin{eqnarray*}
SO(2)=\left\{ M_{2\times 2},\;\; det M=1,\;\; MM^{T}=1\right\},
\end{eqnarray*}
 qui  est un cas particulier  de $SO(n)$
\begin{eqnarray*}
 SO(n)=\left\{ M_{n\times n},\;\;det M=1,\;\;
MM^{T}=1_{n\times n}\right\},
\end{eqnarray*}
admet  une  repr\'esentation de dimension 2
\begin{eqnarray*}
g_{\theta }=\left(
\begin{array}{cc}
\cos \theta  & \sin \theta  \\
-\sin \theta  & \cos \theta
\end{array}%
\right)
\end{eqnarray*}
o\`u  $g_{\left( \theta =0\right) }=e=I_{2\times 2}$ et  $\left(
g_{\theta }\right) ^{-1}=g_{-\theta }$.
\end{enumerate}

\subsection{Repr\'esentations irr\'eductibles et caract\`eres}
Soient $D_{1}(g)$ et $D_{2}(g)$ deux repr\'{e}sentations de $G$ de
m\^{e}me dimension $n$. S'il existe une matrice $A$ (inversible)
telle que  $ D_{2}(g)=A^{-1}D_{1}(g)A$, alors  $D_{1}(g)$ et
$D_{2}(g)$ \ sont dites repr\'esentations \'equivalentes.\\
\\
{\bf Remarque}: On peut consid\'erer  $A$ comme une matrice de
passage agissant sur des bases ($\{\alpha_i\}$ et  ($\{\beta_i\}$)
d'un espace de repr\'esentation $ E$ comme suit $ A: \left\{ \alpha
_{i}\right\} \to \left\{ \beta _{i}\right\}$.\\
\\
{\bf D\'{e}finition}:  On appele caract\`{e}re  d'une
repr\'esentation l'ensemble des nombres
\begin{eqnarray*}
\chi(g) =Tr (D(g)).
\end{eqnarray*}
 Dans une base $\left\{ \mid i\rangle \right\} $ de $E$, si $ D(g)$
\begin{eqnarray*}
D(g)_{ij}\ \ =\langle i\mid D(g)\mid j\rangle.
\end{eqnarray*} alors  on
\'ecrit son caract\`ere par les traces des matrices
\begin{eqnarray*}
\chi(g)  =\sum_iD_{ii}(g)=\sum_i\langle i\mid D(g)\mid i\rangle.
 \end{eqnarray*}
{\bf Remarques}:
\begin{enumerate}
\item Pour une repr\'{e}sentation donne\'{e},  les caract\`{e}res de  tous les \'{e}l\'{e}ments
d'une m\^{e}me classe sont \'{e}gaux. En effet,  si on a deux
\'el\'ement conjug\'{e}s alors  il existe un $g\in G$ tel que
$g_{1}=gg_{2}g^{-1}$,  et donc on a  $D(g_{1})=D(gg_{2}g^{-1})$. En
effet, nous avons
 \begin{eqnarray*}
D(g_{1})=D(g)D(g_{2})D(g)^{-1}.
 \end{eqnarray*}
Si on calcule la trace, on trouve
 \begin{eqnarray*}
TrD(g_{1})&=&Tr(D(g)D(g_{2})D(g)^{-1})=Tr\left(
D(g)D(g^{-1})D(g_{2})\right)=Tr\left( D\left( g_{2}\right) \right)
\\
&=&\chi(g_1)=\chi(g_2).
 \end{eqnarray*}
\item  Deux repr\'{e}sentations \'{e}quivalentes ont m\^{e}me
caract\`{e}res $\chi(g)$ pour tous les \'{e}l\'{e}ments de $G$.
Soient $D_1$ et $D_2$ deux repr\'{e}sentations \'{e}quivalentes
\begin{eqnarray*}
 D_{2}\left( g\right) =  A^{-1}D_{1}\left( g\right)A
 \end{eqnarray*}
En effet, soit $g$ un \'el\'ement de $g$.  On a
 \begin{eqnarray*}
\chi_2(g) &=&Tr D_{2}\left( g\right) = Tr A^{-1}D_{1}\left( g\right)
A\\
&=&Tr A^{-1}AD_{1}\left( g\right)\\
&=&Tr D_{1}\left( g\right)\\
&=& \chi_1(g).
 \end{eqnarray*}
 \item Le caract\`ere de  l'\'el\'ement neutre (identit\'e)  du groupe,
donne  la dimension de la repr\'esentation
 \begin{eqnarray*}
 \chi(e)= dim\;D.
  \end{eqnarray*}
\end{enumerate}
Consid\'erons par exemple le cas du groupe $Z_2$, constitu\'e des
\'el\'ements $e$ et  $g$ tel que $g.g=e$. On connait les
repr\'esentations de  $Z_2$. Il y a une repr\'esentation    triviale
o\`u  tout \'el\'ement correspond \`a  la valeur 1:
 \begin{eqnarray*}
 D(e)=1,\qquad D(g)=1
  \end{eqnarray*}
par cons\'equent
\begin{eqnarray*}
 \chi(e)=1,\qquad \chi(g)=1.
 \end{eqnarray*}
Pour la  deuxi\`eme repr\'esentation de dimension 1 (non triviale),
on a
\begin{eqnarray*}
 \chi(e)=1,\qquad \chi(g)=-1.
 \end{eqnarray*}
Il y aussi une  repr\'esentation de dimension deux
 \begin{eqnarray*}
D_2(e)&=& \left(
\begin{array}{cc}
1 & 0 \\
0 & 1%
\end{array}%
\right) \rightarrow \chi\left( e\right) =2=dim\;D_2\\
 D_2(g)&=&\left(
\begin{array}{cc}
1 & 0 \\
0 & -1%
\end{array}%
\right) \rightarrow \chi\left( g\right) =0.
 \end{eqnarray*}

\subsubsection{Repr\'esentations  r\'{e}ductibles (irr\'{e}dictibles)}
La somme (directe) de deux repr\'{e}sentations $D_{1}\left( g\right) $ et $%
D_{2}\left( g\right) $ d'un m\^{e}me groupe $G$ de dimension
respectives $ n_{1}$ et $n_{2}$ est donne\'{e} par la matrice
$n\times n$ $\left( n=n_{1}+n_{2}\right)$:
\begin{eqnarray*}
D(g)=D_{1}(g)\oplus D_{2}(g)=\left(
\begin{array}{cc}
D_{1}(g) & 0 \\
0 & D_{2}(g)
\end{array}
\right).
\end{eqnarray*}
Dans ce cas, les caract\`{e}res sont les sommes \ \ \ \
\begin{eqnarray*}
\chi\left( D(g)\right) =\chi_{1}\left( g\right) +\chi_{2}\left(
g\right).
\end{eqnarray*}
{\bf  D\'{e}finition}: Une repr\'{e}sentation $D(g)$ est dite r\'{e}ductible si  elle est \'{e}%
quivalente \`{a} une repr\'{e}sentation qui est une somme des
repr\'{e}sentations du m\^{e}me groupe. Autrement,  il faut \
trouver une matrice $A$ telle  que
\begin{eqnarray*}
{D(g)}=A^{-1}D(g)A=\left(
\begin{array}{cc}
D_{1}(g) & 0 \\
0 & D_{2}(g)%
\end{array}%
\right).
\end{eqnarray*}
Dans le cas contraire, elle est irr\'eductible.\\
\\
{\bf Remarque}:  Les repr\'{e}sentations irr\'eductibles d'un groupe
ab\'{e}lien sont toutes de dimension 1.
\section{Repr\'{e}sentation des groupes finis}
Dans cette partie, nous allons nous int\'eresser aux
repr\'esentations irr\'eductibles des groupes finis.\\
\\
{\bf Lemme de Schur}:\\
  Si une matrice $A$  commute avec toutes les
matrices d'une repr\'esentation irr\'eeductible  (de dimension $n$)
\begin{eqnarray*}
 AD(g)=D(g)A\qquad
g\in G,
\end{eqnarray*}
alors $A=\lambda I_{n\times n}$.

\subsubsection{Relation d'orthogonalit\'e:}
Consid\'erons    un groupe fini  $G$ d'ordre $n$.  Soient
$D^{(\alpha )}(g)$ des repr\'esentations irr\'eductibles et
$n_{\alpha }$ leurs dimensions. Soit $\chi^{\alpha }(g)$   le
caract\'{e}re de $D^{\alpha }(g)$. Les matrices $D^{(\alpha) }$
satisfont les propri\'et\'es d'orthogonalit\'e suivantes
\begin{eqnarray*}
 \frac{1}{n}{
\sum_g }D_{ab}^{(\alpha) }\left( g\right) D_{a^{\prime }b^{\prime
}}^{(\alpha^{\prime })}\left( g\right) ^{\ast }=\frac{1}{n_{\alpha
}}\delta _{\alpha \alpha ^{\prime }}\delta _{aa^{\prime }}\delta
_{bb^{\prime }}
\end{eqnarray*}
et leurs caract\`eres satisfont aussi
\begin{eqnarray*}
\frac{1}{n}{%
\sum_g }\chi^{(\alpha) }\left( g\right) \chi^{(\alpha ^{\prime
})}\left( g\right) ^{\ast }=\delta _{\alpha \alpha ^{\prime }}.
\end{eqnarray*}
{\bf Exemple}: Le groupe  cyclique  $Z _{2}=\{e,g\}$.\\ Il y a deux
repr\'esentations irr\'eductibles non \'equivalentes
\begin{table}[th]
\begin{center}
\begin{tabular}{|c|c|c|}\hline
& e & g \\ \hline $D^{\alpha_1}$ & 1 & 1\\ \hline
$D^{\alpha_2}$  & 1 & -1%
\\ \hline
\end{tabular}
\end{center}
\end{table}

La propri\'et\'e d'orthogonalit\'e  pour $Z _{2}$  s'\'ecrit comme
suit
\begin{eqnarray*} \frac{1}{2}\left[ D^{\alpha _{1}}\left(
e\right) D^{\alpha _{2}}\left( e\right) +D^{\alpha _{1}}(g)D^{\alpha
_{2}}\left( g\right) \right] =0\\
\frac{1}{2}\left[ D^{\alpha _{1}}\left( e\right) D^{\alpha
_{1}}\left( e\right) +D^{\alpha _{1}}(g)D^{\alpha _{1}}\left(
g\right) \right] =1.
\end{eqnarray*}
{\bf Th\'{e}or\`eme}: Les dimension $n_{\alpha }$ satisfont
\begin{eqnarray*}
n=\sum_\alpha n_{\alpha }^{2}.
\end{eqnarray*}
{\bf Proposition }: Si $G$ est un groupe  ab\'elien   d'ordre $n$
alors  $n=1^{2}+1^{2}+1^{2}+...+1^{2}$.\\
{\bf Exemple}: Pour le groupe $Z_{2}$, on a $2=1^{2}+1^{2}$.
\section{Tables de caract\`eres}
Dans cette partie,  nous \'etudions la th\'eorie des caract\`eres.
On va  construire une table donnant l'ensemble des caract\`eres d'un
groupe $G$ fini. En effet, l'ensemble des caract\`eres
irr\'eductibles de $G$ forme une base orthonormale. Il se trouve
que,  qu'il y a donc exactement autant de caract\`eres
irr\'eductibles (repr\'esentations irr\'eductibles
non-isomorphes) de $G$ que de classes de conjugaison dans $G$.\\
A titre illustratif, nous donnons  quelques exemples.
\subsubsection{Exemple 1:} Consid\'erons un groupe $G$  qui contient 4
\'{e}l\'{e}ments $G=\left\{ e,C_{2},\sigma _{1},\sigma _{2}\right\}
$. On va donner la  table de multiplication \`a travers l'action de
$G$ sur un vecteur \`{a} 3 dimensions $ \left(\begin{array}{c}
x \\
y \\
z
\end{array}%
\right)$. En effet, nous avons les actions suivantes
\begin{eqnarray*}
e. \left(
\begin{array}{c}
x \\
y \\
z%
\end{array}%
\right)  \longrightarrow \left(
\begin{array}{c}
x \\
y \\
z%
\end{array}%
\right),\qquad C_{2}. \left(
\begin{array}{c}
x \\
y \\
z%
\end{array}%
\right)  \longrightarrow \ \left(
\begin{array}{c}
-x \\
-y \\
z%
\end{array}%
\right) \\
\sigma _{1}. \left(
\begin{array}{c}
x \\
y \\
z%
\end{array}%
\right)  \longrightarrow \left(
\begin{array}{c}
x \\
-y \\
z%
\end{array}%
\right), \qquad \sigma _{2}. \left(
\begin{array}{c}
x \\
y \\
z%
\end{array}%
\right) \longrightarrow \left(
\begin{array}{c}
-x \\
y \\
z%
\end{array}%
\right).
\end{eqnarray*}
La table de multiplication  de ce groupe est donn\'ee par

\begin{center}
\begin{tabular}{|c|c|c|c|c|}
\hline & $e$ & $C_{2}$ & $\sigma _{1}$ & $\sigma _{2}$ \\ \hline $e$
& $e$ & $C_{2}$ & $\sigma _{1}$ & $\sigma _{2}$ \\ \hline $C_{2}$ &
$C_{2}$ & $e$ & $\sigma _{2}$ & $\sigma _{1}$ \\ \hline $\sigma
_{1}$ & $\sigma _{1}$ & $\sigma _{2}$ & $e$ & $C_{2}$ \\ \hline
$\sigma _{2}$ & $\sigma _{2}$ & $\sigma _{1}$ & $C_{2}$ & $e$ \\
\hline
\end{tabular}
\end{center}
On voit que  la table de multplication est sym\'etrique, donc le
groupe est ab\'elien.   Le groupe $G$  poss\`ede 4
repr\'{e}sentations irr\'eductibles car c'est son nombre de classes
de conjugaison.  En particulier, on peut remarquer \'egalement  que
\begin{eqnarray*}
 4=1+1+1+1.
\end{eqnarray*}
 Par cons\'equent  $G$  poss\`ede  4 caract\`eres.   Il est facile de voir que les sous groupes de $G$ sont de type
$Z_2$
\begin{eqnarray*}
\left\{ e, \sigma _{1}\right\},\;\;\left\{ e, \sigma _{2}\right\}
,\;\;\left\{ e, C_{1}\right\}
\end{eqnarray*}
avec la table de multiplication suivante
\begin{center}
\begin{tabular}{|c|c|c|}\hline
& $e$ & $\sigma _{1}(\sigma _{2}, C_2)$ \\ \hline $e$ & $e$ &
$\sigma _{1}(\sigma _{2}, C_2)$
\\ \hline
$\sigma _{1}(\sigma _{2}, C_2)$  & $\sigma _{1}(\sigma _{2}, C_2)$  & $e$ \\
\hline
\end{tabular}
\end{center}
 Pour construire la table des caract\`eres de $G$,  on va  utiliser  la table de caract\`eres  de
 $Z_{2}$. Pour le moment, on a la table de caract\'{e}res suivante:
 \begin{center}
\begin{tabular}{|c|c|c|c|c|}\hline
& $e$ & $C_{2}$ & $\sigma _{1}$ & $\sigma _{2}$ \\ \hline $\chi
_{1}^{1}$ & 1 & 1 & 1 & 1 \\ \hline $\chi _{2}^{1}$ & 1 & $a_1$ &
$a_2$& $a_3$ \\ \hline $\chi _{3}^{1}$ & 1 & $b_1$ & $b_2$ & $b_3$ \\
\hline $\chi _{4}^{1}$ & 1 & $c_1$ & $c_2$ & $c_3$ \\ \hline
\end{tabular}
\end{center}
Pour avoir $a_i$,   $b_i$ et $c_i$ il suffit d'utiliser
l'orthogonalit\'e des colonnes de la table de caract\`eres.  Il faut
r\'esoudre les \'{e}quations suivantes
\begin{eqnarray*}
\left\langle \chi_{1}^{1},\chi_{2}^{1}\right\rangle =\left\langle
\chi_{1}^{1},\chi_{3}^{1}\right\rangle =\left\langle \chi
_{1}^{1},\chi_{4}^{1}\right\rangle =0
\end{eqnarray*}
En effet, on a
\begin{eqnarray*}
1+\chi_{i}^{1}(C_{2})+\chi_{i}^{1}(\sigma _{1})+\chi _{i}^{1}(\sigma
_{2})=0,\qquad i=1,2,3.
\end{eqnarray*}

Pour trouver la solution,  on remarque tout d'abord que  $a_i$,
$b_i$  et $c_i$  sont \'egaux \`a  1 ou -1.  En utilisant les sous
groupes  de type $Z_2$ et  le caract\`ere non trivial,  on trouve
que exactement deux des $a_i$ $(b_i,c_i)$ valent -1. Donc finalement
la table de caract\`eres de $G$ est  donn\'ee  par
\begin{center}
\begin{tabular}{|c|c|c|c|c|}\hline
& $e$ & $C_{2}$ & $\sigma _{1}$ & $\sigma _{2}$ \\ \hline $\chi
_{1}^{1}$ & 1 & 1 & 1 & 1 \\ \hline $\chi_{2}^{1}$ & 1 & -1 & 1 & -1
\\ \hline $\chi_{3}^{1}$ & 1 & -1 & -1 & 1
\\ \hline $\chi_{4}^{1}$ & 1 & 1 & -1 & -1\\ \hline
\end{tabular}
\end{center}

\subsubsection {Exemple 2: groupe $S_3$.} Commen\c{c}ons d'abord par
rappler que $S_3$ est d'ordre 6 ($\mid S_{3}\mid =6=3!$). $S_{3}$
contient les \'el\'ements
\begin{center}
 $\ \ \ \ \ \left(
\begin{array}{ccc}
1 & 2 & 3 \\
1 & 2 & 3%
\end{array}%
\right), \left(
\begin{array}{ccc}
1 & 2 & 3 \\
1 & 3 & 2%
\end{array}%
\right), \left(
\begin{array}{ccc}
1 & 2 & 3 \\
3 & 2 & 1%
\end{array}%
\right), \left(
\begin{array}{ccc}
1 & 2 & 3 \\
2 & 1 & 3%
\end{array}%
\right), \left(
\begin{array}{ccc}
1 & 2 & 3 \\
3 & 1 & 2%
\end{array}%
\right), \left(
\begin{array}{ccc}
1 & 2 & 3 \\
2 & 3 & 1%
\end{array}%
\right) $. \end{center}
Les dimensions des repr\'{e}sentations irreductibles
satisfont  la relation suivante
\begin{eqnarray*}
6=\sum_\alpha n_{\alpha}^{2}
\end{eqnarray*}
Pour $S_3$,  on  a 3 trois classes   de conjugaison diff\'{e}rents
qui sont comme suit
\begin{center}
$\left\{ \left(
\begin{array}{ccc}
1 & 2 & 3 \\
1 & 2 & 3%
\end{array}%
\right) \right\} ,\left\{ \left(
\begin{array}{ccc}
1 & 2 & 3 \\
1 & 3 & 2%
\end{array}%
\right) \left(
\begin{array}{ccc}
1 & 2 & 3 \\
3 & 2 & 1%
\end{array}%
\right) \left(
\begin{array}{ccc}
1 & 2 & 3 \\
2 & 1 & 3%
\end{array}%
\right) \right\} ,\left\{ \left(
\begin{array}{ccc}
1 & 2 & 3 \\
3 & 1 & 2%
\end{array}%
\right) \left(
\begin{array}{ccc}
1 & 2 & 3 \\
2 & 3 & 1%
\end{array}%
\right) \right\}. $
\end{center}
Ces  trois classes de conjugaison correspondent  aux partitions
suivantes $ (.)(.)(.),(.)(..),(...)$ qui  poss\`edent $1,3,2$
\'el\'ements resepectivement. $S_3$  a  3 caract\`eres car c'est son
nombre de classes de conjugaison. En effet, nous avons
\begin{eqnarray*}
6=\sum_{\alpha=1}^3 n_{\alpha}^{2}=1^{2}+1^{2}+2^{2},\qquad
n_{1}=1,\; n_{2}=1,\;n_{3}=2.
\end{eqnarray*} On a deux repr\'{e}sentations de
dimension 1 et une repr\'{e}sentation de dimension 2. On obtient
donc pour le moment  la table de caract\`eres (7.1)  avec
$\chi_3=( 2, a,b)$ o\`u $a$, $b$ sont deux coefficients \`a
d\'eterminer.
\begin{table}[th]
\begin{center}
\begin{tabular}{|c|c|c|c|}\hline
&(.)(.)(.) & (.)(..) & (...) \\ \hline $\chi_1$& 1 & 1 & 1\\ \hline
$\chi_2$ & 1& -1& 1
\\ \hline $\chi_3$ & 2& $a$&$b$
\\ \hline
\end{tabular}
\caption{La table de multiplication} \label{t:two}
\end{center}
\end{table}

 Pour avoir $a$ et $b $ il suffit d'utiliser
l'orthogonalit\'e des colonnes de la table de caract\`eres. En
effet, l'orthogonalit\'e des caract\`eres donne tout de suite
\begin{eqnarray*}
\langle \chi_{1}\mid \chi_{2}\rangle &=&\frac{1}{6}\left[
2+3a+2b\right]
=0\\
\langle \chi_{1}\mid \chi_{3}\rangle &=&\frac{1}{6}\left[
2-3a+2b\right] =0,
 \end{eqnarray*}  ceci nous donne que   $a = 0$ et $b =
-1$. Donc la table de caract\`eres de $S_3$  prend la forme suivante

\begin{table}[th]
\begin{center}
\begin{tabular}{|c|c|c|c|}\hline
&(.)(.)(.) & (.)(..) & (...) \\ \hline $\chi_1$& 1 & 1 & 1\\ \hline
$\chi_2$ & 1& -1& 1
\\ \hline $\chi_2$ & 2& 0&-1
\\ \hline
\end{tabular}
\end{center}
\end{table}

\bigskip
La repr\'esentation irr\'eductible de dimension 2  de $S_3$  est
donn\'ee par
\begin{eqnarray*}
\left(
\begin{array}{ccc}
1 & 2 & 3 \\
1 & 2 & 3%
\end{array}%
\right)&\equiv& \left(
\begin{array}{cc}
1 & 0 \\
0 & 1%
\end{array}%
\right), \qquad
 \left(
\begin{array}{ccc}
1 & 2 & 3 \\
1 & 3 & 2%
\end{array}%
\right)\equiv \frac{1}{2}\left(
\begin{array}{cc}
1 & \sqrt{3} \\
-\sqrt{3} & -1%
\end{array}%
\right), \\
\left(
\begin{array}{ccc}
1 & 2 & 3 \\
3 & 2 & 1%
\end{array}
\right)&\equiv& \frac{1}{2}\left(
\begin{array}{cc}
1 & \sqrt{3} \\
\sqrt{3} & -1%
\end{array}%
\right),\qquad  \left(
\begin{array}{ccc}
1 & 2 & 3 \\
2 & 1 & 3%
\end{array}%
\right)\equiv \left(
\begin{array}{cc}
-1 & 0 \\
0 & 1%
\end{array}%
\right)\\
 \left(
\begin{array}{ccc}
1 & 2 & 3 \\
1 & 3 & 2%
\end{array}%
\right)&\equiv& \frac{1}{2}\left(
\begin{array}{cc}
-1 & \sqrt{3} \\
-\sqrt{3} & -1%
\end{array}%
\right),\qquad
 \left(
\begin{array}{ccc}
1 & 2 & 3 \\
1 & 3 & 2%
\end{array}%
\right) \equiv \frac{1}{2}\left(
\begin{array}{cc}
-1 & -\sqrt{3} \\
\sqrt{3} & -1%
\end{array}%
\right).
 \end{eqnarray*}
\subsubsection {Exemple 3: groupe $S_4$: Exercice}

\chapter{   Groupes de Lie}
\pagestyle{myheadings}
\markboth{\underline{\centerline{\textit{\small{ Groupes de
Lie}}}}}{\underline{\centerline{\textit{\small{ Groupes de Lie}}}}}
Les groupes de Lie $G$ d\'{e}pendent de param\`{e}tres continus. En
g\'en\'erale,  chaque \'{e}l\'{e}ment $g$ est d\'etermin\'e  par
\begin{eqnarray*}
g=g(\theta _{1},...,\theta _{n}),\quad  et \;\; e=f(0,0,...,0)
\end{eqnarray*}
o\`{u} $n$ est la dimension  de $G$.\\
Commen\c{c}ons par un exemple particuli\`{e}rment simple.

\section{Le groupe de Lie $SO(2)$} $SO(2)$ est le
groupe des rotations en deux dimensions. Il d\'epend d'un seul
 param\`{e}tre continu \`{a} savoir $\theta$ o\`u $0\leq \theta \leq
2\pi$. Il contient une repr\'{e}sentation \emph{\ unitaire } de
dimension deux\footnote{\emph{ rep. unitaire : $ g\longrightarrow $
op\'{e}rateur unitaire. Une fois on a une base  $ g \longrightarrow
$ Matrice}}.
 Cette repr\'{e}sentation bi-dimensionelle est dite
aussi repr\'esentation fondamentale du groupe $SO (2)$. Dans ce cas,
on a
\begin{eqnarray*}
D(\theta )=\left(
\begin{array}{cc}
\cos \theta  & -\sin \theta  \\
\sin \theta  & \cos \theta
\end{array}%
\right),
\end{eqnarray*}
avec
\begin{eqnarray*}
D(\theta )D(\theta )&=&D(\theta )D^{T}(g)=1\\
\det D(\theta )&=&1.
\end{eqnarray*}
La loi de composition de deux rotations  est donn\'ee par
\begin{eqnarray*}
\ D(\theta )D(\theta ^{\prime })\ =D(\theta +\theta ^{\prime }\;
mod\; 2\pi ).
\end{eqnarray*}
{\bf Remarques}
\begin{enumerate}
\item $SO(2)$  est un groupe ab\'{e}lien. En effet, on a
\begin{eqnarray*}
D(\theta )D(\theta ^{\prime })\ =D(\theta ^{\prime })D(\theta
)=D(\theta +\theta ^{\prime }{mod}\;2\pi )
\end{eqnarray*}
\item  Utilisant le fait que
$\theta =n\frac{\theta }{n}=\frac{\theta }{n}+\frac{\theta
}{n}+...+\frac{ \theta }{n}$,
 on peut d\'{e}monter que
 \begin{eqnarray*}
D(\theta )=\left[ D(\frac{\theta }{n})\right] ^{n}.
\end{eqnarray*}
\end{enumerate}
 Si on prend $n\rightarrow \infty $, on trouve des rotations
(transformations) infinit\'{e}simales d'angle $\frac{\theta }{n}$.
Dans ce cas, on a
\begin{eqnarray*}
D(\frac{\theta }{n})&=&\left(
\begin{array}{cc}
\cos \frac{\theta }{n} & -\sin \frac{\theta }{n} \\
\sin \frac{\theta }{n} & \cos \frac{\theta }{n}%
\end{array}%
\right) \\
 &=&\left(
\begin{array}{cc}
1+... & -\frac{\theta }{n}+... \\
\frac{\theta }{n}+... & 1+...%
\end{array}%
\right)
\\&=&\left(
\begin{array}{cc}
1 & 0 \\
0 & 1%
\end{array}%
\right) +i\frac{\theta }{n}\left(
\begin{array}{cc}
0 & i \\
-i & 0%
\end{array}%
\right)\\&=&I_{2\times 2}+i\frac{\theta }{n}T+...
\end{eqnarray*}
o\`{u} $T$ est une matrice hermitienne
\begin{eqnarray*}
T=\left(
\begin{array}{cc}
0 & i \\
-i & 0%
\end{array}%
\right) =T^{+}.
\end{eqnarray*}
Calculons la puissance $n$ (exercice). En effet,  on trouve
\begin{eqnarray*}
D(\theta )= \sum_{k=0}^{\infty }\frac{\left( i\theta T \right)^{k}}{
k!}=\exp \left( i\theta T\right),  \qquad T^{2}=1,\;\; T^{2k+1}=T.
\end{eqnarray*}
Si on prend $T = 0$ , on trouve $ D(\theta )=1$ qui est la
repr\'{e}sentation triviale  de dimension 1. Si on pose  $T= m$
\begin{eqnarray*}
D(\theta )=e^{im\theta }
\end{eqnarray*}
o\`{u} $m$
 est un entier,   on trouve une famille des rep\'{e}sentations non trivales
de dimension 1 ( o\`u $m = 0$ est la repr\'{e}sentation triviale).\\
Plus g\'{e}n\'eralement, o\`{u} nous avons plusieurs  param\'{e}tres
par exemple le groupe  $(SO(n))$,  on a la formule suivante:
\begin{eqnarray*}
\ \ D(g)=1+i\sum_{a=1}^{\dim G}\theta _{a}T^{a}+...
\end{eqnarray*}
o\`{u} les matrices $T^{a}$  satisfont les relations suivantes
\begin{eqnarray*}
\ \left[ T^{a},T^{b}\right] =\sum_{c=1}^{\dim G}f_{c}^{ab}t^{c}.
\end{eqnarray*}

\section{Le groupe SU(2)}
Le groupe SU(2) est form\'{e} par des matrices unitaires ($2\times
2$) et de d\'eterminant 1. La forme g\'{e}nerale d'un \'el\'ement de
$SU(2)$ est donn\'{e}e par
\begin{eqnarray*}
D\left( g\right) =\left(
\begin{array}{cc}
a & b \\
-b^{\ast } & a^{\ast }%
\end{array}%
\right)
\end{eqnarray*}
 o\`{u} $a$ et $b$ sont deux complexes. La condition de unitarit\'{e} et le determinant = 1 est assur\'{e}e par
\begin{eqnarray*}
\mid a\mid ^{2}+\mid b\mid ^{2}=1.
\end{eqnarray*}
La dimension de $SU(2)$ est 3($ 2\times 2$, deux  complexes - une
condition  r\'{e}elle=3.  Si on \'{e}crit les deux nombres complexes
sous la forme suivante
\begin{eqnarray*}
 a&=&a_{1}+ia_{2}
\\
b&=&b_{1}+ib_{2},
\end{eqnarray*}
la  condition $\mid a\mid ^{2}+\mid b\mid ^{2}=1$ devient
\begin{eqnarray*}
a_{1}^{2}+a_{2}^{2}+b_{1}^{2}+b_{2}^{2}=1
\end{eqnarray*}
et montre que
le groupe $SU(2)$ est comme une sph\`{e}re $S^{3}$ dans l'espace $R^{4}$.\\
\\
{\bf  Remarque}: Le groupe $SU(2)$ est un groupe compact, ($SU(2)\sim S^{3}$).\\
\\
Le groupe $SU(2)$ est parametris\'{e} par
\begin{eqnarray*}
a&=&\cos \theta \cos \varphi e^{i\psi }\\
b&=&\sin \theta \cos \varphi +i\sin \varphi.
\end{eqnarray*}
La rep\'{e}sentation de dimension 2 de $SU(2)$ qui d\'{e}finit le groupe lui-m%
\^{e}me est
\begin{eqnarray*} D(\theta
,\varphi ,\psi )=\left(
\begin{array}{cc}
\cos \theta \cos \varphi e^{i\psi } & \sin \theta \cos \varphi
+i\sin
\varphi  \\
-\sin \theta \cos \varphi +i\sin \varphi  & \cos \theta \cos \varphi
e^{-i\psi }%
\end{array}%
\right)
\end{eqnarray*}
o\`u  $\theta \in \left[ 0,2\pi \right] $ et  $(\psi,\varphi)   \in
\left[ \frac{-\pi }{2},\frac{\pi }{2}\right] $. Pour la suite, nous
utiliserons le changement
\begin{eqnarray*}
 \theta _{1}=\varphi \\
 \theta _{2}=\theta \\
 \theta_{3}=\psi
\end{eqnarray*}
 ce qui nous  permet de trouver la forme suivante
\begin{eqnarray*}
  D(\theta _{1},\theta _{2},\theta _{3})=\left(
\begin{array}{cc}
\cos \theta _{1}\cos \theta _{2}e^{i\theta _{3}} & \sin \theta
_{1}\cos
\theta _{2}+i\sin \theta _{1} \\
-\sin \theta _{2}\cos \theta _{1}+i\sin \theta _{1} & \cos \theta
_{1}\cos
\theta _{2}e^{-i\theta _{3}}%
\end{array}%
\right).
\end{eqnarray*}
Maintenant, on  peut \'{e}crire
\begin{eqnarray*}
D\left( \theta _{1},\theta _{2},\theta _{3}\right) =I_{2\times
2}+i\sum_{a=1}^{3}{\theta _{a}T^{a}}+...
\end{eqnarray*}
o\`u
\begin{eqnarray*}
T^{1}=\left(
\begin{array}{cc}
1 & 0 \\
0 & 1%
\end{array}%
\right), \quad T^{2}=\left(
\begin{array}{cc}
0 & -i \\
i & 0%
\end{array}%
\right), \quad T^{3}=\left(
\begin{array}{cc}
1 & 0 \\
0 & -1%
\end{array}%
\right).
\end{eqnarray*}
Les matrices $T^{a}$ ( matrices de Pauli) v\'erifient les relations
de commutation suivantes
\begin{eqnarray*}
 \left[ T^{a},T^{b}\right] =2i\varepsilon_{abc}T^{c}.
\end{eqnarray*}

\section{Le groupe $SO(3)$}
Ce sont les rotations en trois dimensions  qui  transforment un vecteur \`{a}
trois \ composantes  \`{a} un autre vecteur \`{a} trois composantes comme suit
\begin{eqnarray*}
R: \left(
\begin{array}{c}
V_{1} \\
V_{2} \\
V_{3}%
\end{array}%
\right) {\rightarrow }  \left(
\begin{array}{c}
V_{1}^{\prime } \\
V_{2}^{\prime } \\
V_{3}^{\prime }%
\end{array}%
\right)
\end{eqnarray*}
o\`u
\begin{eqnarray*}
\ \ R=\left(
\begin{array}{ccc}
&  &  \\
&  &  \\
&  &
\end{array}%
\right) _{3\times 3}
\end{eqnarray*}
o\`u  $V^{\prime} =RV$.  La matrice $R$  d\'{e}pend en g\'{e}nerale
de trois param\`{e}tres (angles d'Euler).  En utilisant les
rotatations infinit\'{e}simales, il est facile de trouver que,
\begin{eqnarray*}
R\left( \theta _{1}\right) R\left( \theta _{2}\right) R\left( \theta
_{3}\right) =1+i\theta _{1}J_{1}+i\theta _{2}J_{2}+i\theta _{3}J\
\end{eqnarray*}
o\`u
\begin{eqnarray*}
J_{1}=\left(
\begin{array}{ccc}
0 & 0 & 0 \\
0 & 0 & -i \\
0 & i & 0%
\end{array}%
\right), \qquad J_{2}=\left(
\begin{array}{ccc}
0 & 0 & i \\
0 & 0 & 0 \\
-i & 0 & 0%
\end{array}%
\right), \qquad J_{1}=\left(
\begin{array}{ccc}
0 & -i & 0 \\
i & 0 & 0 \\
0 & 0 & 0%
\end{array}%
\right).
\end{eqnarray*}
On peut  les exprimer  aussi comme suit
\begin{eqnarray*}
\left( J_{k}\right) _{ij}=i\varepsilon_{kij}
\end{eqnarray*}
\`a l'aide du tenseur $\varepsilon_{kij}$.  Ces  trois matrices
$J_i, \;(i=1,2,3)$  satisfont les relations de commutation suivantes
\begin{eqnarray*}
\left[ J_{a},J_{b}\right] =i\varepsilon_{abc}J_{c}.
\end{eqnarray*}

\section{Repr\'{e}sentations de $SO(3)$ et $SU(2)$}
Les  g\'en\'erateurs infinit\'esimaux $J_i$   de SO(3) et $T_i$ de
SU(2) satisfont les  m\^emes relations de commutation. Ces
g\'en\'erateurs sont reli\'es par la formule suivante
\begin{eqnarray*}
J_i=\frac{T_i}{2}.
\end{eqnarray*}
 Les repr\'esentations de $SU(2)$  et  $SO(3)$ sont donne\'{e}s par
\begin{eqnarray*}
D\left( \theta \right) =\exp \left( {\sum_{i=1}^{3} \theta _{i}}%
J_{i}\right).
\end{eqnarray*}
\subsubsection{Repr\'{e}sentation de $SU(2)$} Il est utile
d'utiliser une base  form\'ee  par les vecteurs
\begin{eqnarray*}
\{J_{+},J_{-}, J_{3}\}
\end{eqnarray*}
o\`u $J_{\pm }=J_{1}\pm iJ_{2}$. Ces g\'{e}n\'{e}rateurs $J_{1}$,
$J_{2}$ et $J_{3}$ sont hermitiennes  et $J_{\pm }$ ne le sont plus.
En effet, nous avons
\begin{eqnarray*}
J_{3}^{+}=J_{3},\qquad J_{\pm }^{+}=J_{\mp}.
\end{eqnarray*}
Il est  facile  de calculer
\begin{eqnarray*}
\left[ J_{3},J_{\pm }\right] &=&\pm J_{\pm }\\
\left[ J_{+},J_{-}\right] &=&2J_{3},
\end{eqnarray*}
o\`u $J_{3}$ et \ $J_{\pm }$ sont des op\'erateurs agissant dans un
espace vectoriel $V$.  Pour donner une repr\'esentation, il faut
chercher leurs actions sur $V$.  Supposons que l'espace vectoriel
$V$ admet une repr\'esentation irr\'eductible de demention finie. Le
but est
de  construire une base de $V$.\\
\\
{\bf Remarque}: Si $\mid m\rangle $\ est un \'etat propre de
$J_{3}$\ de valeur propre $m$, $ J_{3}\mid m\rangle =m\mid m\rangle
$, alors $J_{\pm }\mid m\rangle $
 sont \'{e}galement des \^{e}tats propres de $J_{3}$ de valeur propre $m\pm
 1$. \\
En effet, on a
\begin{eqnarray*}
J_{3}J_{\pm }\mid m\rangle &=&\left( \left[ J_{3},J_{\pm }\right]
+J_{\pm }J_{3}\right) \mid m\rangle \\
 &=&\left( \pm J_{\pm }+mJ_{\pm
}\right) \mid m\rangle \\&=&\left( m\pm 1\right) J_{\pm }\mid
m\rangle.
\end{eqnarray*}
Consid\'{e}rons l'op\'erateur suivant
\begin{eqnarray*}
 J^{2}&=&J_{1}^{2}+J_{2}^{2}+J_{3}^{2}.
\end{eqnarray*}
On peut v\'erifier que  $J^{2}$ commute avec  tous les $J_i$
\begin{eqnarray*}
\left[ J^{2},J_{i}\right] =0,\qquad i=1,2,3.
\end{eqnarray*}
Il est imm\'ediat de calculer
\begin{eqnarray*}
\ \left[ J^{2},J_{1}\right] &=&\left[
J_{3}^{2}+J_{2}^{2},J_{1}\right] \\&=&\left[ J_{2}^{2},J_{1}\right]
+\left[ J_{3}^{2},J_{1}\right]
\\ &=&J_{2}\left[ J_{2},J_{1}\right] +\left[ J_{2},J_{1}%
\right] \bigskip J_{2}+J_{3}\left[ J_{3},J_{1}\right] +\left[ J_{3},J_{1}%
\right]  \\&=&-iJ_{2}J_{3}-iJ_{3}J_{2}+iJ_{3}J_{2}+iJ_{2}J_{3}\\&=&0.
\end{eqnarray*}
$J^{2}$ est dit l'op\'{e}rateur de Casimir.  On peut l'\'{e}crire
aussi comme  suit \begin{eqnarray*}
 J^{2}&=&J_{1}^{2}+J_{2}^{2}+J_{3}^{2}\\
 &=&J_{-}J_{+}+J_{3}^{2}+J_{3}\\&=&J_{+}J_{-}+J_{3}^{2}-J_{3}
\end{eqnarray*}
puisque  $J_{+}J_{-}=J_{1}^{2}+J_{2}^{2}+i\left[ J_{2},J_{1}\right]
=J_{1}^{2}+J_{2}^{2}+J_{3}$.\\ $J^{2}$ commute avec les trois
g\'{e}n\'{e}rateurs, alors $J^{2}=\lambda 1$.\\
\\
Soit $\mid j\rangle $ un \'etat propre de $J_3$. Nous obtenons la forme suivante
\begin{eqnarray*}
J^{2}\mid j\rangle =J_{-}J_{+}\mid j\rangle +J_{3}^{2}\mid j\rangle
+J_{3}\mid j\rangle =J_{-}J_{+}\mid j\rangle +\left( j^{2}+j\right)
\mid j\rangle.
\end{eqnarray*}
Supposons que $j$  est  une  valeur propre maximale autrement
$J_{+}\mid j\rangle =0$, alors
\begin{eqnarray*}
J^{2}\mid j\rangle =j\left( j+1\right) \mid j\rangle.
\end{eqnarray*}
 Finalement, nous donnons  le  r\'esultat suivant:
 \begin{itemize}
\item Pour chaque dimension de $V$, l'alg\`ebre  de Lie $su(2)$ et  $so(3)$ poss\`edent
des  repr\'{e}sentations irr\'eductibles.
\item La dimension de la repr\'{e}sentation est $2j+1$ avec $%
j=0,\frac{1}{2},1,\frac{3}{2},...$
\item On a les relations suivantes
\begin{eqnarray*}
J_{+}\mid m\rangle &=&\sqrt{j(j+1)-m(m+1)}\mid m+1\rangle
\\
J_{3}\mid m\rangle &=&m\mid m\rangle, \qquad -j \leq m\leq j
\\
J_{-} \mid m\rangle &=&\sqrt{j(j+1)-m(m-1)}\mid m-1\rangle
\end{eqnarray*}
 \end{itemize}
{\bf Exemple:}  $j=\frac{1}{2}$. On obtient la repr\'esentation de
dimenion 2 suivante
\begin{eqnarray*}
J_{+}=\left(
\begin{array}{cc}
0 & 1 \\
0 & 0%
\end{array}%
\right),\qquad J_{-}=\left(
\begin{array}{cc}
0 & 0 \\
1 & 0%
\end{array}%
\right), \qquad J_{3}=\left(
\begin{array}{cc}
\frac{1}{2} & 0 \\
0 & -\frac{1}{2}%
\end{array}%
\right).
\end{eqnarray*}

\section{Les groupes $SU(n)$ et $SO(n)$}
\subsection{ Le groupe $SU(n)$}
 Le groupe $SU(n)$  poss\`ede une importance particuli\`ere en physique des particules.  Par exemple,   le groupe unitaire $U(1)$ qui est isomophe \`a $SO(2)$  est le groupe de jauge de l'\'electromagn\'etisme, et $SU(2)$ est le groupe associ\'e \`a  l'interaction faible.   Le groupe  $SU(3)$ correspond  \`a  l'interaction forte (mod\`eles  des quarks).\\
 $SU(n)$ est le groupe des matrices  $n \times n$ unitaires \`a  coefficients complexes    et de d\'eterminant 1:
\begin{eqnarray*}
SU(n)=\{M\in {\cal M}_n(C):\;\;  det M=1,\quad MM^+=I_{n\times n}\}.
\end{eqnarray*}
Le $SU(n)$  est un groupe de Lie r\'eel de dimension
$n^2-1$(compact). L'alg\`ebre de Lie correspondant \`a  $SU(n)$ est
not\'ee $A_{n-1}=su(n)$.
\subsection{ Le groupe $SO(n)$}
Le groupe sp\'ecial orthogonal $SO(n, R)$ est le groupe des matrices
orthogonales $n\times n$ \`a  coefficients dans  $R$:
\begin{eqnarray*}
SO(n)=\{M\in {\cal M}_n(R):\;\;  det M=1,\quad MM^t=I_{n\times n}\}
\end{eqnarray*}
$SO(n)$ peut \^etre consid\'er\'e  comme le groupe des rotations
dans l'espace $ R^n$.  Il est compact et de dimension
$\frac{n(n-1)}{2}$.

La deuxi\`eme partie de ces notes  sera trait\'ee prochainement.

\end{document}